\documentclass[prd,twocolumn,amsmath,amssymb,floatfix,superscriptaddress,nofootinbib,preprintnumbers]{revtex4-1}
\pdfoutput=1
\usepackage{color}
\usepackage{graphicx}
\usepackage{tabularx}
\usepackage[dvipsnames, table]{xcolor}
\usepackage{xspace}
\usepackage[justification=centerlast, format=plain, labelfont=bf]{caption}
\usepackage{subcaption}
\usepackage{float}
\usepackage{verbatim}
\usepackage{amsmath}
\usepackage{amssymb}
\usepackage{url}
\usepackage{bbold}
\usepackage{slashed}
\usepackage{array}
\usepackage{multirow}
\usepackage{threeparttable}
\usepackage{paralist}
\usepackage{xspace}
\usepackage{upgreek}
\usepackage{lipsum}
\usepackage{ulem}

\usepackage[colorlinks]{hyperref}
\newcolumntype{C}{>{\centering\arraybackslash}X}

\newcommand{\taur}{{\tau_{\mathrm{reio}}}}

    \newcommand{\be}{\begin{equation}}
  \newcommand{\ee}{\end{equation}}
    \newcommand{\ba}{\begin{align}}
  \newcommand{\ea}{\end{align}}

\newcommand{\du}{{\rm d}}

\newcommand{\eV}{\textrm{eV}}
\newcommand{\Planck}{{\it Planck }}

\newcommand{\limr}{\textcolor{black}{LiMR}}
\newcommand{\limrs}{\textcolor{black}{LiMRs}}

\begin{document}

\title{Finding eV-scale Light Relics with Cosmological Observables}
\author{Nicholas DePorzio}
\affiliation{Department of Physics, Harvard University, Cambridge, MA 02138, USA}
\author{Weishuang Linda Xu}
\affiliation{Department of Physics, Harvard University, Cambridge, MA 02138, USA}
\author{Julian B.~Mu\~noz}
\affiliation{Department of Physics, Harvard University, Cambridge, MA 02138, USA}
\author{Cora Dvorkin}
\affiliation{Department of Physics, Harvard University, Cambridge, MA 02138, USA}

\begin{abstract}
Cosmological data provide a powerful tool in the search for physics beyond the Standard Model (SM).
An interesting target are light relics, new degrees of freedom which decoupled from the SM while relativistic.
Nearly massless relics contribute to the radiation energy budget, and are commonly parametrized as variations in the effective number $N_{\rm eff}$ of neutrino species.
Additionally, relics with masses greater than $10^{-4}$ eV become non-relativistic before today, and thus behave as matter instead of radiation.
This leaves an imprint in the clustering of the large-scale structure of the universe, as light relics have important streaming motions, mirroring the case of massive neutrinos.
Here we forecast how well current and upcoming cosmological surveys can probe light massive relics (LiMRs). We consider minimal extensions to the SM by both fermionic and bosonic relic degrees of freedom. 
By combining current and upcoming cosmic-microwave-background and large-scale-structure surveys, we forecast the significance at which each LiMR, with different masses and temperatures, can be detected. 
We find that a very large coverage of parameter space will be attainable by upcoming experiments, opening the possibility of exploring uncharted territory for new physics beyond the SM. 
\end{abstract}

\maketitle

%%%%%%%%%%%%%%%%%%%%%%%%%%%%%%%%%%%%%%%%%%%%%%%%%%%%
\section{Introduction}
 
The nature of the dark sector is one of the major puzzles of fundamental physics, integral to the understanding of our universe across almost every epoch. 
Searches for the composition of the dark sector and, more broadly, of physics beyond the Standard Model (SM), take place at different energy scales, and use data ranging from particle colliders to astrophysical and cosmological surveys. 
The interactions of the dark sector with the SM are central to many of these searches. 
Yet, the small energies and interaction cross-sections expected in many models often result in low experimental sensitivity to new physics. In contrast, by exploring the entropic effects of new dark-sector physics, cosmological data is in an exciting position to make robust discoveries.

Numerous extensions of the SM happen to posit the existence of light, feebly interacting particles, including axions and axion-like particles~\cite{Peccei:1977hh,Weinberg:2013aya,Svrcek:2006yi,Arvanitaki:2009fg}, dark photons~\cite{Abel:2008ai,Beranek:2013yqa,ArkaniHamed:2008qp,Essig:2009nc}, and light fermions~\cite{Cheung:2007ut, Goldberg:1986nk,Feldman:2007wj}.  
One broad category are \textit{light relics}, stable particles which were in thermal contact with the SM in the early universe and decoupled while relativistic. Consequently, their cosmic abundance was frozen and survived until $z=0$. The quintessential example within the SM are neutrinos, but they need not be the only light relics to populate our universe. 
Different proposed new light relics include a fourth, sterile neutrino, whose existence is suggested by different anomalous experimental results~\cite{Fukuda:2000np,Mention:2011rk,Aguilar-Arevalo:2013pmq} (see Ref. \cite{Dentler:2018sju} for a recent review); as well as the gravitino, the supersymmetric partner of the graviton~\cite{Benakli:2017whb}.

New relics that are sufficiently light will manifest as dark radiation, and can be searched for through their effect on the cosmic microwave background (CMB) anisotropies~\cite{Bashinsky:2003tk,Hou:2011ec,Baumann:2015rya}, typically parametrized by the effective number of neutrino species, $N_{\rm eff}$ (which is 3.045 in the standard cosmological model~\cite{Mangano:2005cc,deSalas:2016ztq, Akita:2020szl}). 
Massive relics can, on the other hand, become non-relativistic at some point in cosmic history, and behave as other components of matter in the Universe thereafter. However, their decoupling while relativistic gives these relics significant streaming motion, which sets a scale below which they cannot cluster, thus altering the large-scale structure (LSS) of our universe.  
This has allowed cosmology to set the leading constraints on neutrino mass, at $\Sigma m_\nu{<}0.26$ eV ($95\%$ C.L.), assuming standard cosmology~\cite{Aghanim:2018eyx}. 
In this work we will search for new Light---but Massive---Relics ({\limrs}) using cosmological observables.

Cosmological data from near-future surveys are expected to provide exquisite measurements of the distribution of matter in our universe.
{\limrs} that have become non-relativistic before $z=0$ (with masses $m_X\gtrsim$ $10^{-3}$ eV), will impact that distribution by behaving as hot dark matter~\cite{Boyarsky:2008xj,Banerjee:2016suz,Baur:2017stq,Dodelson:2016wal,Baumann:2017lmt}.
In addition to the relic mass, two relevant parameters determine the relic abundance.
The first is their number $g_X$ of degrees of freedom. 
The second is their temperature $T_X^{(0)}$ today.
Due to comoving-entropy conservation, any relic that was in equilibrium with the SM in the early universe ought to have $T_X^{(0)} \geq 0.91$ K. This minimum temperature gives rise to different values of $\Delta N_{\rm eff}$ for each type of relic~\cite{Brust:2013ova}: 0.027 for scalars ($g_{X}=1$), 0.047 for Weyl fermions ($g_X=2$), 0.054 for massless gauge bosons ($g_X=2$), 
and 0.095 for Dirac fermions ($g_X=4$).  
In addition, relics with masses in the eV-scale will become non-relativistic before $z=0$, leaving an imprint in the form of suppressed matter fluctuations. Here we forecast how well eV-scale {\limrs} can be observed by joint CMB and LSS surveys.

This paper is structured as follows. In Section~\ref{sec:effects} we briefly review light relics and their effects on cosmological observables. In Section~\ref{sec:methods} we detail the datasets we consider, which we employ in Section~\ref{sec:results} to forecast constraints on {\limrs}  within the mass range $10^{-2}$ eV- $10^1$ eV. We conclude in Section~\ref{sec:conclusions}.

%%%%%%%%%%%%%%%%%%%%%%%%%%%%%%%%%%%%%%%%%%%%%%%%%%%%
\section{Light relics and their effect on cosmological observables}
\label{sec:effects}

We begin with an overview of the physics of light relics and their effects on cosmological observables.  
A {\limr} $X$ is characterized by its present-day temperature $T_X^{(0)}$ and mass $m_X$, as well as its statistics, bosonic or fermionic, and its number $g_X$ of degrees of freedom.  
The present-day temperature of a light relic (massive or not) is set by the time at which it decouples from the SM thermal bath, which is found as 
\be
T_X^{(0)} = \left( \frac{g_{*S}^{(0)}}{\,\,g^{(\rm dec)}_{*S}} \right)^{1/3}T_{\gamma}^{(0)}, 
\ee
where $g_{*S}^{(0/\rm dec)}$ denotes the entropy degrees of freedom in the universe today/when the relic decoupled, and $T_{\gamma}^{(0)}=2.725 $ K is the present-day temperature of the photon bath. In this way, the conservation of comoving entropy provides a minimal light relic temperature assuming the SM with no additional degrees of freedom (other than the relic), 
\be
T_{X}^{(0)} \gtrsim \left( \frac{3.91}{106.75} \right)^{1/3}T_{\gamma}^{(0)} \approx 0.91 \; \mathrm{K}, 
\ee
where just after the electroweak phase transition we have $g^{(\rm dec)}_{*s}=106.75$ encompasses all the known degrees of freedom of the Standard Model, and the present-day value of $g_{*s}^{(0)}=3.91$ includes photons and decoupled, cooler neutrinos. As an example, the SM (active) neutrinos have $T_\nu^{(0)}=1.95$ K, as they decoupled just prior to electron-positron annihilation where $g_{*s}^{(\rm dec,\nu)}=10.75$.  Note that the baryonic and cold-dark matter (CDM) contributions are negligible, given their exponentially suppressed abundance. 

In contrast, light relics decoupled while relativistic, and so are cosmologically abundant, with number densities comparable to that of photons or neutrinos. For instance, a Weyl fermion decoupling as early as possible (with minimal present-day temperature 0.91 K) will have a number density today of 11 cm$^{-3}$, and a vector boson that decouples just before e$^+$e$^-$ annihilation (with a temperature today of 1.95 K, as neutrinos) will have a present-day number density of 150 cm$^{-3}$. Thus, the contribution of light relics to the cosmic energy budget can be significant. 

It is often enlightening to describe the cosmological effects of other relics in relation to those of neutrinos, given their common origin as light relics.  As advanced in the introduction, relics in the early universe (while $T_X\gg m_X$) behave as radiation, and their cosmological impact while relativistic can be encapsulated in the number of effective neutrinos, $N_{\rm eff}$, defined with respect to their contribution to the radiation energy density,
\begin{align}
\rho_{\rm rad}(z) &= \frac{\pi^2}{30} \left(  \sum_{\rm bosons} g_b T_b^4(z) + \frac{7}{8} \sum_{\rm fermions} g_f T_f^4(z) \right) \nonumber \\
& \equiv \frac{\pi^2}{30} \left(2 T_{\gamma}^4(z) + \frac{7}{4} N_{\rm eff} T_{\nu}^4(z) \right),
\label{eq:ComovingEntropy}
\end{align}
where $T_{\gamma/\nu}(z)$ is the temperature of photons and neutrinos at redshift $z$, $g_b$/$g_f$ are the degrees of freedom, and $T_b$/$T_f$ are the temperatures of each boson/fermion, respectively. 

Introducing an entropically significant light relic will generate a contribution to Eq.~\eqref{eq:ComovingEntropy} of $(\pi^2 / 30) g_X T_X^4$ for bosonic species, or $7/8$ times that for fermionic species. We can then describe any departure from the predicted value of $N_{\textrm{eff}}^{ \Lambda\textrm{CDM}}=3.045$ in the standard $\Lambda$CDM model by the quantity $\Delta N_{\rm eff}$, given by
\be
\label{eq:Neff_relc=ic}
\Delta N_{\rm eff} = c_1^\gamma \left(\frac{g_{X}}{g_{\nu}}\right) \left(\frac{T_{X}^{(0)}}{T_{\nu}^{(0)}}\right)^4,
\ee
in terms of the neutrino parameters $g_\nu=2$ and $T_{\nu}^{(0)}=1.95$ K.
The factor $c_1=8/7$ accounts for the difference between the Bose-Einstein ($\gamma=1$) and Fermi-Dirac ($\gamma=0$) distributions.

\begin{figure}[h!]
    \centering
    \includegraphics[width = \linewidth]{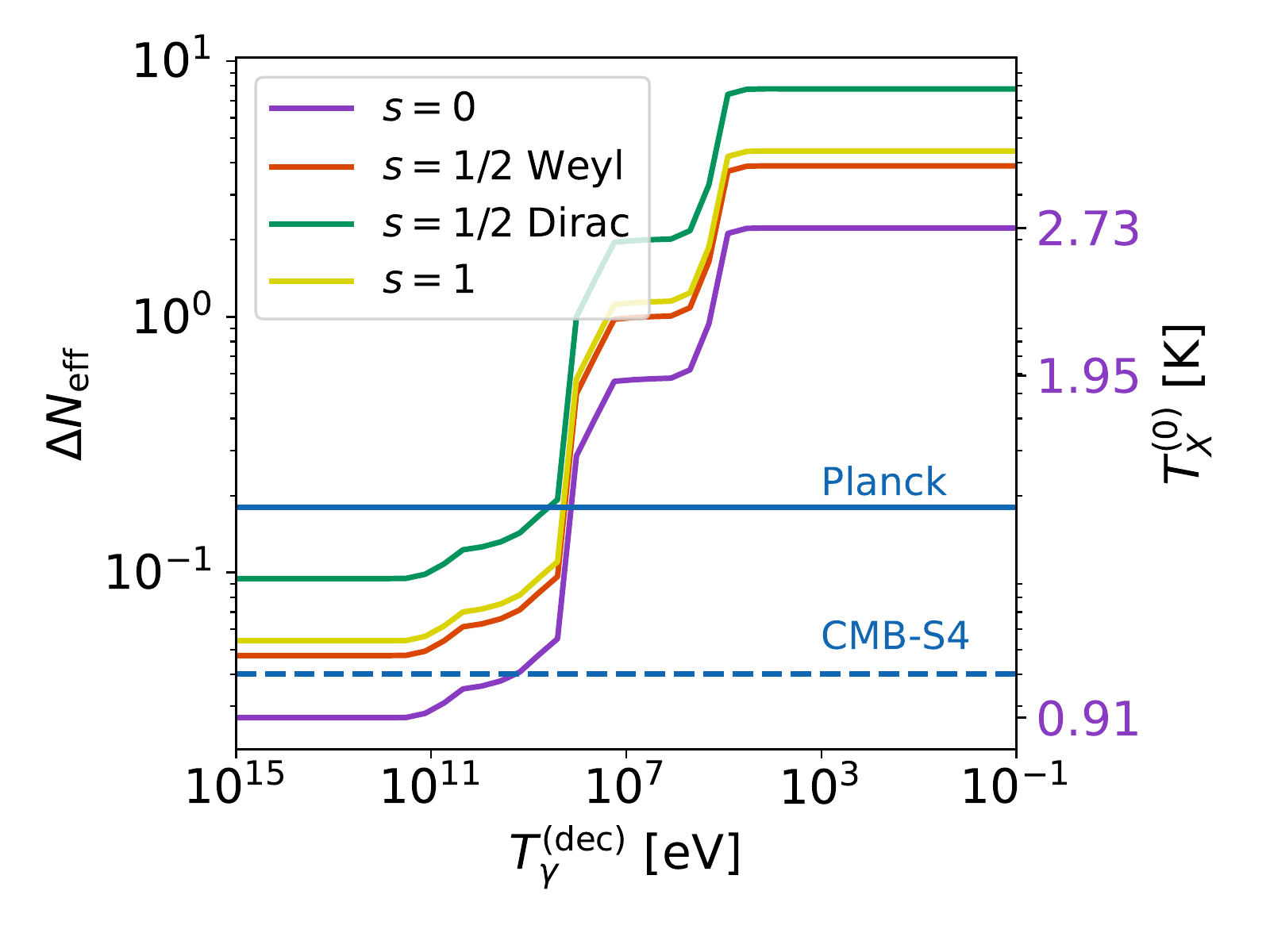}
    \caption{Cosmic evolution of $\Delta N_{\textrm{eff}}$ due to a light relic that decoupled when the universe had a temperature $T_\gamma^{(\rm dec)}$.
    We assume four different types of relics with spin $s$, as described in the text, and show the $68\%$ C.L. constraints achieved by \Planck as a horizontal solid line, and the forecast by {CMB-S4} in dashed lines.
    The right vertical axis shows what the temperature of the relic would be at $z=0$, following the violet (lowest) curve plotted for $s=0$.
    Note that these constraints only apply to relics with $m_X \approx 0.1$ eV or lighter.}
    \label{fig:Neffplot}
\end{figure}

This discussion is encapsulated in Fig.~\ref{fig:Neffplot}, showing the relation between the present-day relic temperature to the time of relic decoupling, and its corresponding contribution to $N_{\rm eff}$. Note that the present-day temperature of a relic for fixed decoupling epoch does not depend on particle species, but its contribution to radiation energy does. 

\captionsetup[figure]{justification=centerlast}

Current limits on $\Delta N_{\rm eff}$ arise primarily from observables at two epochs. The first is recombination. Measurements of radiation at recombination are sensitive to relics lighter than  $\sim 0.1$ eV. The \Planck 2018 analysis reports a measurement of $N_{\rm eff}=2.99^{+0.34}_{-0.33}$ (TT+TE+EE+lowE+lensing+BAO) at 95\% C.L.~\cite{Aghanim:2018eyx}. The proposed CMB-Stage 4 (CMB-S4) experiment is expected to refine this measurement to the $\sigma(N_{\textrm{eff}})=0.03$ level \cite{Abazajian:2016yjj}. 
The second is the Helium abundance, from where we can infer the number of relativistic species present during big bang nucleosynthesis (BBN). The 68\% C.L.~measurement during that era is $N_{\textrm{eff}}=2.85 \pm 0.28$ \cite{Cyburt:2015mya}, 
which is valid for all relics lighter than $m_X\lesssim 10^6$ eV.
Note that this does not affect dark matter (DM) produced via the freeze-in mechanism, as it can contribute negligibly to $N_{\rm eff}$ \cite{Cheung:2011nn,Dvorkin:2019zdi}. 

In this work we consider detection prospects for four types of {\limrs}: scalars, vectors, and both Dirac and Weyl fermions.
We study relics with eV-scale masses, $10^{-2}\,{\rm eV} \leq m_X \leq 10^1\,\eV$, such that they all behave as matter at $z=0$, with the highest mass candidates constituting up to ${\sim}10\%$ of DM abundance. Finally, we also consider a range of temperatures, bounded by $T^{(0)}_X \geq 0.91$ K from below. 
Our maximum temperature is informed by  the constraint $\Delta N_{\rm eff} \leq 0.36$ from {\it Planck}, corresponding to a single additional species of Weyl fermion at $T^{(0)}_{X} \leq 1.5 \; \mathrm{K}$.
This bound could be further improved by combining with BBN measurements of e.g. D/H ratios~\cite{Fields:2019pfx}, Lyman-$\alpha$ forest flux power spectrum data~\cite{Rossi:2014nea, Palanque-Delabrouille:2015pga}, as well as Baryon Acoustic Oscillations (BAO) and galaxy power spectrum measurements~\cite{Alam:2016hwk, Zarrouk:2018vwy, Baumann:2017gkg}.

%%%%%%%%%%%%%%%%%%%%%%%%%%%%%%%%%%%%%%%%%%%%%%%%%%%%
\subsection*{Effect on the LSS of the Universe}

{\limrs} can become non-relativistic at some point in cosmic history, and comprise a fraction of DM at $z=0$.
Unlike CDM, which is expected to compose the majority of the matter sector, {\limrs} have significant thermal motions, even if non-relativistic. Thus, these relics will stream away from structures below their free-streaming scale, which during matter domination is given by~\cite{AliHaimoud:2012vj,Bird:2018all} 

\be
k_{\rm fs}  = \frac{0.08}{\sqrt{1+z}}\left( \dfrac{m_X}{0.1\rm eV}\right)\left( \dfrac{T_{X}^{(0)}}{T_\nu^{(0)}}\right)^{-1}  h\,\rm Mpc^{-1}.
\label{eq:kfs}
\ee

Throughout this section we assume a Weyl fermionic relic, and we will relax this assumption later.
This presents another way of searching for {\limrs}: through their effect on the matter fluctuations. {\limrs} produce a suppression in the matter power spectrum at scales smaller than $k_{\rm fs}$, which we discuss below. The size of this suppression depends on the present abundance of the {\limr}, 
which (if non-relativistic) is given by
\be
\Omega_X h^2  = \dfrac{m_{X}}{93.14 \, {\rm eV}} \frac{g_X}{g_\nu}   \left( \dfrac{T_{X}^{(0)}}{T_\nu^{(0)}}\right)^3.
\label{eq:omegaX}
\ee
From Eq.~\eqref{eq:omegaX} we see that there is a maximum allowed particle mass, found by saturating the observed DM abundance $\Omega_{\rm cdm}h^2=0.12$ \cite{Aghanim:2018eyx}. 
For a relic temperature $T_X^{(0)}{\approx}1.5$ K, this is $m_X{\approx}10$ eV. Additionally, in this work we are interested in the relics that become non-relativistic before today.
Thus, the mass range we will study encompasses

\be
{\rm 10^{-2}\, eV} \leq m_{X} \leq 10^1 \,\rm eV.
\ee

{\limrs} produce a suppression in matter fluctuations, similar to neutrinos, due to two reasons. The first is simply that the light relic does not cluster at small scales, and its fluctuation $\delta_X$ at small-scale roughly follows 
$\delta_X = \left(k/k_{\rm fs}\right)^{-2} \delta_m
$
with respect to the matter overdensity $\delta_m$.
The second is that the absence of relic fluctuations at small scales slows down the growth of CDM (and baryon) overdensities.
Together, these two factors produce a suppression of roughly $(1-14 f_X)$ in the matter power spectrum~\cite{Lesgourgues:2006nd}, where $f_X$ is the fraction of matter that is composed of the \limr  $\,X$.
This suppression is less pronounced for relics that stay relativistic for longer, which yields the well-known result of $(1-8f_\nu)$ for neutrinos comprising a fraction $f_\nu$ of matter, as neutrinos only become non-relativistic during matter domination.
These numbers are for illustration purposes only, and in all cases we find the full effect of {\limrs} on the cosmological observables using the publicly available software {\tt CLASS}~\cite{Blas:2011rf}. 
Nevertheless, they provide intuition about the physical effect of such a relic. While the mechanism that produces the suppression is the same as for neutrino masses, the free-streaming scale $k_{\rm fs}$ for a {\limr} is not fully determined by its mass (or abundance), as their temperature today is unknown. 
Relics that are still relativistic at $z=0$ (with $m_{\rm X} \lesssim  10^{-3}$ eV) will have never collapsed into structures and thus their observable effects can be fully included into $\Delta N_{\rm eff}$. 
In practice, this is the case for {\limrs} with masses below $\sim 0.1$ eV, as we will show, so we will use our results for a $10^{-2}$ eV relic for lighter masses.

To study {\limrs}, the relevant observables are the fluctuations of baryons and cold dark matter, as only those will gravitationally bind to form the visible structures we observe as galaxies, the relics being too light to cluster (see, however, Ref.~\cite{LoVerde:2013lta}).
The power spectrum of baryonic plus cold dark-matter fluctuations is modeled by
\be
P_{\rm cb} (k) = P_\zeta(k) \Big( f_b T_b(k) + f_{c} T_{c}(k) \Big)^2,
\ee
where $P_\zeta$ is the primordial power spectrum,  the transfer functions $T_{b}$ and $T_{c}$ are found using {\tt CLASS}~\cite{Blas:2011rf}, 
and the fractional abundances are defined by
\be
f_{b/c} \equiv \frac{\omega_{b/c}}{\omega_b + \omega_{c}},
\ee
where $\omega_b$ and $\omega_c$ are the baryon and CDM abundances.

We show the suppression in $P_{\rm cb}$ in Fig.~\ref{fig:SigmaBossFine} (upper panel) for a fermion with $m_X=0.02$ eV and $T_X=0.91$ K, for degrees of freedom $g_X=2,$ 3 and 4.
In all cases the high-$k$ power is more suppressed, as expected.
Increasing the abundance of the {\limr}, by augmenting $g_X$, produces a more marked suppression, while keeping the shape fixed.
Moreover, increasing the relic abundance produces wiggles at the BAO scale, as the {\limr} both contributes as radiation at recombination and free streams -- like neutrinos -- changing the BAO phase~\cite{Bashinsky:2003tk}.

The suppression of matter fluctuations produces a change in the biasing of galaxies, which has been calculated for both neutrinos and other relics~\cite{LoVerde:2014pxa,Munoz:2018ajr,Chiang:2018laa},
and accounted for in neutrino-mass forecasts in our companion paper \cite{Xu:2020fyg}. This produces a growth in the galaxy power spectrum that partially compensates the relic-induced suppression. 
Here we account for this growth induced scale-dependent bias (GISDB) by multiplying the Lagrangian bias by a $k$-dependent factor
\be
g(k) =  R_L^{\rm \Lambda CDM}(k) R_L^{X}(k)  R_L^{\nu}(k),
\label{eq:g}
\ee
where the functions $R_L^i$ account for different effects, following Ref.~\cite{Munoz:2018ajr}.
First, $R_L^{\rm \Lambda CDM}$ accounts for the step-like change in the growth rate of fluctuations before and after matter-radiation equality,
parametrized  as
\be 
R_L^{\rm \Lambda CDM}(k)=1+\Delta_{\rm \Lambda CDM} \tanh \left(\frac{\alpha k}{k_{\rm eq}}\right),
\ee
where $\Delta_{\Lambda \textrm{CDM}}=4.8\times10^{-3}$ and $\alpha=4$
determine the amplitude and location of the step, given the scale $k_{\rm eq}$ of matter-radiation equality. 
The two other factors account  for the effect of a {\limr} on the matter power spectrum, also taken to be a step-like function
\be
R_L^{i}(k) = 1 + \Delta_{i} \tanh \left( 1+\frac{\ln{q_{i}(k)}}{\Delta q} \right),
\ee 
with an amplitude $\Delta_{i} = 0.6 f_i$ determined by the fraction $f_{i}$ of matter composed of the relic $i$ ($X$ or $\nu$),  width $\Delta q=1.6$,
and where we have defined $q_{i}(k) \equiv 5 k/k_{\textrm{fs,{i}}}$,
given the free-streaming scale $k_{\textrm{fs},i}$ of each {\limr}.

%%%%%%%%%%%%%%%%%%%%%%%%%%%%%%%%%%%%%%%%%%%%%%%%%%%%
\subsection*{Effect on the CMB}

The CMB is sensitive to the presence of {\limrs} in the universe, through their mean energy density \cite{1966ApJ...146..542P,PhysRevD.26.2694} and their perturbations \cite{PhysRevD.69.083002,2013PhRvD..87h3008H}. 
Their additional energy density changes the expansion rate of the universe, which in turn affects the CMB damping tail. Since matter-radiation equality is very well measured through the location of the first acoustic peak, this causes the power spectrum to be suppressed on short-wavelength modes. In addition to this effect, their perturbations cause a change in the amplitude and a shift in the location of the CMB acoustic peaks (for a review of the phase shift in the acoustic peaks in the CMB, see Ref.~\cite{Baumann:2015rya}).

\begin{figure}[h!]
    \centering
    \begin{subfigure}{\linewidth}
    \centering
    \includegraphics[width = \textwidth]{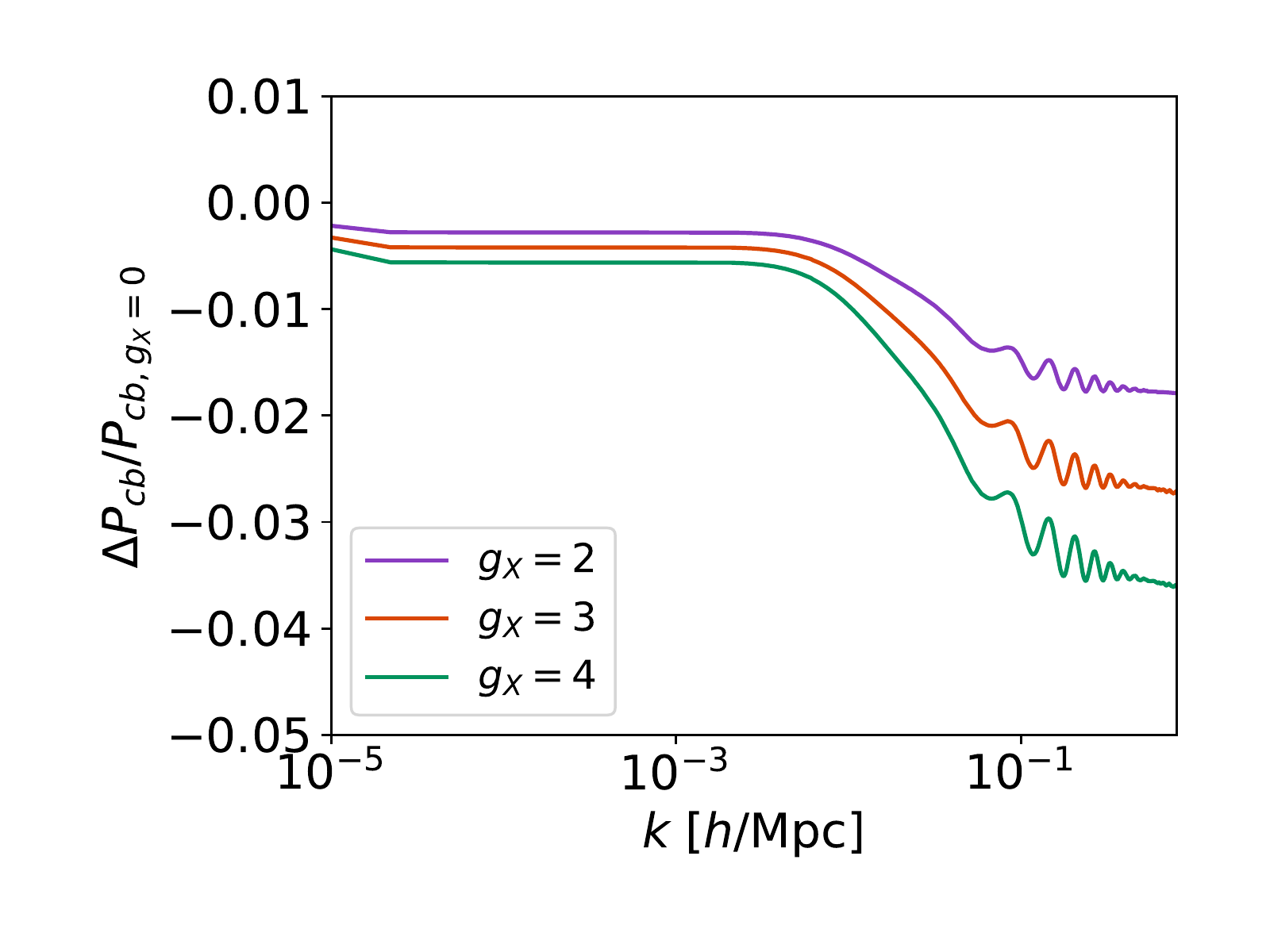}
    \end{subfigure}
    \begin{subfigure}{\linewidth}
    \centering
    \includegraphics[width = \textwidth]{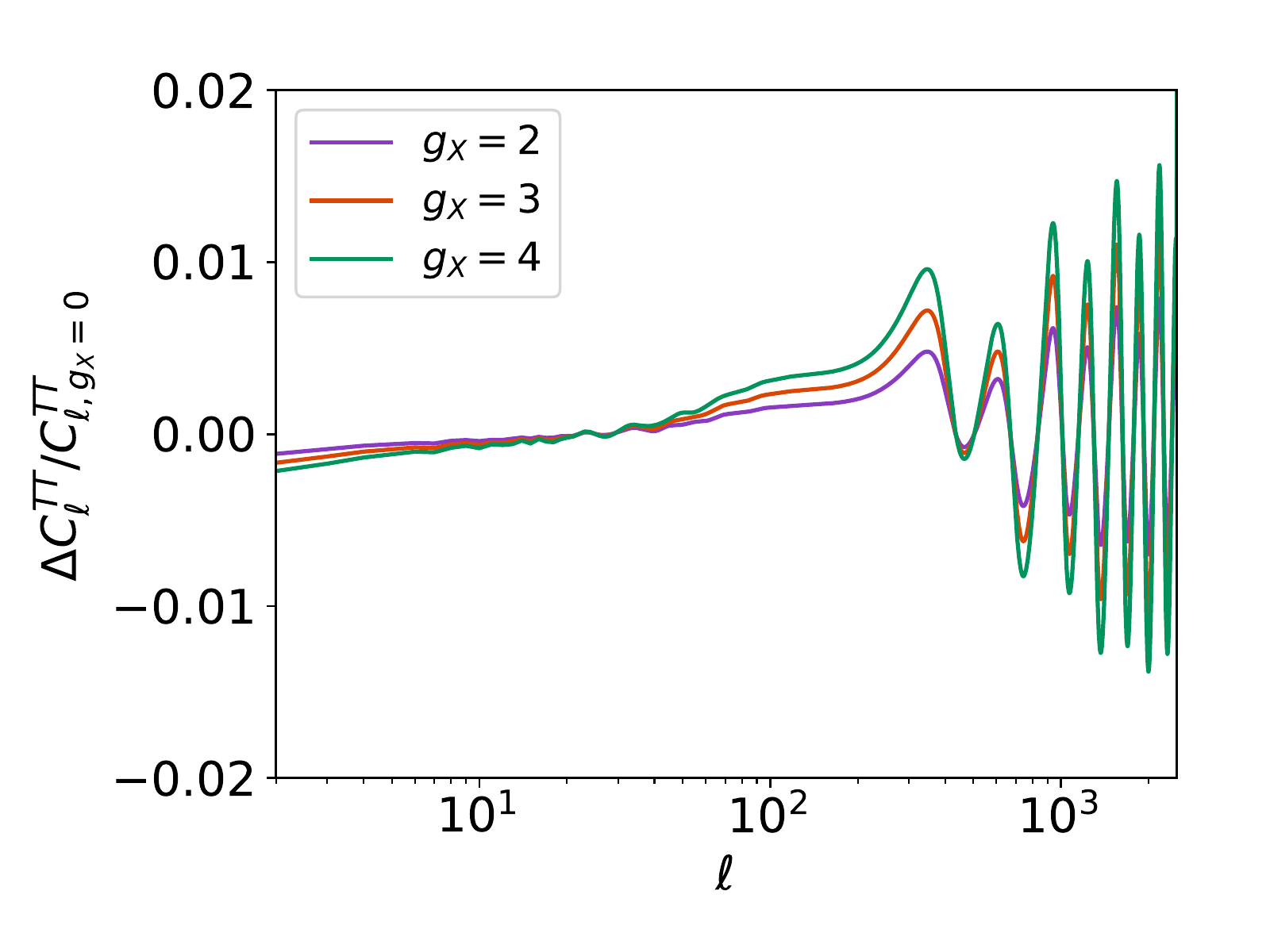}
    \end{subfigure}
    \caption{Effect of introducing a fermion with degrees of freedom $g_X$, temperature $T_X = 0.91$ K and mass $m_X = 0.02$ eV on the CDM+baryon power spectrum (upper panel) and the CMB temperature power spectrum (lower panel). Here all cosmological parameters are fixed when introducing the {\limr} so the fraction of the matter or radiation energy occupied by the LiMR before and after its non-relativistic transition will increase with its abundance. Since the LiMR energy density is not counted in the CDM plus baryon power power spectrum, an increase in LiMR abundance will manifest as an overall suppression to $P_{cb}$. We note that an effective fractional number of degrees of freedom may be achieved as a result of out-of-equilibrium processes.}
    \label{fig:SigmaBossFine}
\end{figure}

We show an example of the effect of a {\limr} on the CMB in Fig.~\ref{fig:SigmaBossFine} -- again for a fermion with $m_X=0.02$ eV and $T_X=0.91$ K, for degrees of freedom $g_X=2,$ 3 and 4. The amplitude and phase shift of the BAO is clearly seen to increase with $g_{\rm X}$.

\captionsetup[figure]{justification=centerlast}

%%%%%%%%%%%%%%%%%%%%%%%%%%%%%%%%%%%%%%%%%%%%%%%%%%%%
\subsection*{Types of Relic}

Throughout this work we will study four major types of {\limrs}, two fermionic and two bosonic, which we now describe.

In the fermionic category, the first type we study are the neutrino-like Weyl fermions, with non-zero mass, spin $s=1/2$, and two degrees of freedom ($g_{X}=2$).
In addition to sterile neutrinos, an intriguing example is the gravitino, the supersymmetric partner of the graviton. While the gravitino has $s=3/2$, only the longitudinal modes couple to the Standard Model and hence behaves equivalently to an $s=1/2$ particle with $g_X=2$. The gravitino is predicted in models of supersymmetric gravity to have a mass in the eV range~\cite{Osato:2016ixc,Martin:1997ns}, within the range relevant to our study.
The second type we tackle are the related Dirac fermions, such as the axino~\cite{Covi:1999ty}, which simply have twice as many degrees of freedom ($g_X=4$).

In the bosonic category we study two types of particles as well: firstly scalars, with only one degree of freedom ($g_X=1$).
A realization of this model could be a Goldstone boson, which can have naturally small masses. 
The second type are spin-1 vectors.
We assume that they have a Stueckelberg mass, as it is technically natural~\cite{Reece:2018zvv} and avoids complications from Higgs mechanisms.
While this relic will be non-relativistic today, its longitudinal mode was decoupled in the early universe (while it was relativistic), and thus only two of the three degrees of freedom were populated. 
Therefore, this relic has $g_X=2$.

Instead of modifying the distribution function for each type of relic, we will take advantage of the fact that any relic, whether bosonic or fermionic, can be recast onto an equivalent Weyl relic (i.e., a neutrino with $g_W=2$), with some temperature $T_W^{\rm eq}$ and mass $m_W^{\rm eq}$~\cite{Munoz:2018ajr,Boyarsky:2008xj}. Justification for this procedure is based on the results of other works which considered the significance of the distribution shapes for different species \cite{Munoz:2018ajr}.
Assuming a relic of temperature $T_X$, with $g_X$ degrees of freedom, the equivalent Weyl relic has
\begin{align}
T_{W}^{\rm eq} &= T_X  \left(g_X/g_{W}\right)^{1/4} c_1^{\gamma/4} \\ 
m_{W}^{\rm eq} &= m_X  \left(g_X/g_{W}\right)^{1/4} c_1^{\gamma/4} c_2^{\gamma},
\end{align}
where we correct for the different distributions of these particles by setting $\gamma=1$ for bosons (and $\gamma=0$ for our base case of fermions as before), with constants $c_1=8/7$ (as in Eq.~\ref{eq:Neff_relc=ic}) and $c_2=7/6$.
Note that our normalization is slightly different from that found in Ref.~\cite{Munoz:2018ajr}, as there fermionic degrees of freedom contributed by 3/2.

%%%%%%%%%%%%%%%%%%%%%%%%%%%%%%%%%%%%%%%%%%%%%%%%%%%%
\section{Methods}
\label{sec:methods}

We now present our forecasting methods.
In this first exploratory work we will follow a Fisher-matrix approach, in order to efficiently explore the 2D parameter space ($T_X^{(0)}, m_X$) of possible {\limrs}.
We encourage the reader to visit Appendix~\ref{sec:AppMCMC} for a comparison against MCMC results.
We will also cover different combinations of datasets.
For the CMB, we will study the current \Planck satellite~\cite{Aghanim:2018eyx} as well as the upcoming ground-based {CMB-S4}~\cite{Abazajian:2019tiv}.
On the galaxy-survey side we will consider the current BOSS~\cite{Dawson_2012}, the ongoing {DESI}~\cite{Aghamousa:2016zmz}, and the upcoming {\it Euclid}~\cite{Amendola:2016saw} surveys. 

%%%%%%%%%%%%%%%%%%%%%%%%%%%%%%%%%%%%%%%%%%%%%%%%%%%%
\subsection{Parameters}

We are interested in forecasting how well different {\limrs} with varied temperatures and masses can be detected.
Therefore, a simple Fisher forecast of the relic mass and temperature, assuming a particular fiducial relic, is insufficient.
Instead, we will find how well {\limrs} of varying mass $m_X$ and temperature $T_X^{(0)}$ can be observed by different experiments.
The parameter we will forecast is $g_X$, the number of degrees of freedom of the {\limr}.\footnote{We note that, while $g_X$ appears to be a fixed quantity for a given relic, e.g. $g_X=1$ for a scalar, changing $g_X$ simply means altering the amount of relic particles (as both $\Delta N_{\rm eff}\propto g_X$ and $\Omega_X \propto g_X$) while keeping their thermal properties identical. That makes $g_X$ a useful variable to forecast.}
Then, $g_X/\sigma(g_X)$ is a good proxy for the significance at which a {\limr} of a particular $m_X$ and $T_X^{(0)}$ can be detected.

In order to properly search for a {\limr} we have to marginalize over the six $\Lambda$CDM parameters.
These include the baryon and cold dark-matter abundances, $\omega_b$ and $\omega_{\rm cdm}$ (with fiducial values of $\omega_b=$ 0.02226 and $\omega_{\rm cdm}=$ 0.1127), the (reduced) Hubble constant $h=0.701$, and the optical depth $\tau_{\rm reio}=$ 0.0598 to reionization.
The last two parameters are the amplitude $A_s$, and tilt $n_s$, of primordial fluctuations, with fiducial values of
$A_s=2.2321 \times 10^{-9}$ and $n_s=$ 0.967. 
In addition, we marginalize over the effect of neutrino masses.
We assume for our fiducial model the existence of three degenerate massive neutrinos, with $\sum m_\nu = 0.06$ eV, and we will report constraints both with and without marginalization over neutrino masses. Unless explicitly stated, no prior will be assumed for these parameters in the Fisher forecasts used to provide parameter constraints. 
For a discussion about the effect of the neutrino hierarchy see Refs.~\cite{Archidiacono:2020dvx, Xu:2020fyg}.

%%%%%%%%%%%%%%%%%%%%%%%%%%%%%%%%%%%%%%%%%%%%%%%%%%%%
\subsection{CMB experiments}

We will model both \Planck and CMB-S4 as having a single effective observing frequency, to avoid marginalizing over foregrounds. 
For \Planck we will use CMB temperature ($T$) and $E$-mode polarization data, covering the range $\ell=[2-2500]$.
We take  noises of $\Delta_T=43 \mu$K-arcmin and $\Delta_E=81 \mu$K-arcmin,
with a $\theta_{\rm FWHM}=5$ arcmin angular resolution.
This well approximates the (more complex) \Planck data likelihood.

For {CMB-S4} we take $\Delta_T=1 \mu$K-arcmin, and $\Delta_E=\sqrt{2}\Delta_T$, with an angular resolution of $\theta_{\rm FWHM}=3$ arcmin.
Additionally, we include lensing data, where we perform iterative delensing of $B$-modes to lower the noise, as in Refs.~\cite{Hirata:2003ka,Okamoto:2003zw}.
All modes cover the range $\ell=[30-5000]$, except for the $TT$ autocorrelation, where we do not go beyond  $\ell=3000$ to avoid foreground contamination~\cite{Abazajian:2016yjj}.
We add a Gaussian prior on the optical depth of reionization of $\sigma(\taur)=0.01$, instead of the $\ell<30$ modes in this case.
This follows the prescription in the {CMB-S4} Science Book~\cite{Abazajian:2016yjj}, as well as our companion paper \cite{Xu:2020fyg} and is the sensitivity reported from the \Planck 2018 results. As such, it serves as a conservative estimate for futuristic surveys, such as {CMB-S4}.~\cite{Aghanim:2018eyx} 

The CMB data will perform two main roles.
First, it will very precisely measure the standard cosmological parameters, breaking many degeneracies in the LSS data.
Second, the CMB is sensitive to the effects of a {\limr} both during recombination and in the matter fluctuations at lower redshifts, through the weak lensing information.

%%%%%%%%%%%%%%%%%%%%%%%%%%%%%%%%%%%%%%%%%%%%%%%%%%%%
\subsection{Galaxy surveys}

For the LSS data we will consider three surveys, all of them spectroscopic. We leave for future work studying the promise of photometric surveys, such as the Vera Rubin Observatory~\cite{Ivezic:2008fe}, and weak-lensing surveys, such as the Dark Energy Survey~\cite{Abbott:2005bi}.

We take the luminous red galaxy (LRG) sample of the Sloan Digital Sky Survey Baryon Oscillation Spectroscopic Survey ({BOSS})~\cite{Dawson_2012}, which will serve as an indication of the power of current data. To showcase the promise of upcoming surveys we study the emission-line galaxy (ELG) sample of the Dark Energy Spectroscopic Instrument ({DESI})~\cite{Aghamousa:2016zmz}, and the more futuristic H$\alpha$-emitters of {\it Euclid}~\cite{Amendola:2016saw}. We restrict our analysis to a single tracer, the most populous for each survey, though more optimistic results are expected for multi-tracer approaches~\cite{Boyle:2017lzt}. The noise per redshift bin for each sample is reported in Table~\ref{tab:ELGs}. We assume sky coverages of 10,000 deg$^2$ for {BOSS}; 14,000 deg$^2$ for {DESI}; and 15,000 deg$^2$ for {\it Euclid}.

As each of these surveys contain distinct tracers, the bias description of each will be somewhat different as well.
Here we follow a simple approach, and parametrize the linear Eulerian bias as
\be
b_1 (k,z) = \left [ 1 + b_L(k,z) + \alpha_{k2} k^2 \right],
\ee
where the $\alpha_{k2}$ term (with a fiducial value of $1\,\rm Mpc^2$) accounts for non-linearities in the bias~\cite{Modi:2016dah}. We emphasize that we do not include the clustering of light relics in this description. We also note that while cold dark matter and baryons may demonstrate different clustering behaviors at small scales, we do not consider such scales in this work and so do not include corrections to the bias that would differentiate the baryon and cold dark matter clustering fields. An additional scale-dependence comes from the aforementioned GISDB effect, which enters in the Lagrangian bias, 
\be
b_L(k,z) = \left[b_0(z) - 1\right] g(k),
\ee
where $g(k)$ is as defined in Eq.~\eqref{eq:g}. The redshift evolution of the bias is encapsulated in the term $b_0 (z)$, which is chosen such that the scale-independent (i.e., $k\to0$) behavior of the Eulerian bias matches with suggestions made elsewhere in the literature \cite{Munoz:2018ajr}. For the ELGs in DESI we match to
\be
b_0(z) = \dfrac{\beta_0}{D(z)},
\ee
where $D(z)$ is the growth factor and $\beta_0 =1$~\cite{Aghamousa:2016zmz}; whereas for the tracers in BOSS and {\it Euclid} we take
\be
b_0(z) = \beta_0 (1+z)^{0.5\beta_1},
\ee
with fiducials $\beta_0 =1.7$ and $\beta_1=1$ as in Ref.~\cite{Sprenger:2018tdb}. We marginalize over the nuisance parameters $ \beta_0$, $\alpha_{k2}$, as well as $\beta_1$ for BOSS and \textit{Euclid}.
We note that a full analysis of the data might require marginalization over the amplitude of the bias at each redshift bin independently, which would however lead to a loss in constraining power.

\begin{table*}[t!]
    \begin{center}
    \vspace{3mm}
        \begin{tabularx}{\textwidth}{ c | CCCCCCCCCC  }
        \toprule
       $z$ & 0.05 & 0.15 & 0.25 & 0.35 & 0.45 & 0.55 & 0.65  & 0.75 & 0.85 & 0.95   \\ 
       \hline \hline 
       \rule{0pt}{4ex} $\displaystyle{\frac{\du N_{LRG}}{\du z \; \du\mathrm{deg}^2}}$ [{BOSS}] & 8 & 50 & 125 & 222 & 332 & 447 & 208 & 30 & 0 & 0 \\
      \cline{2-11}
      \rule{0pt}{4ex} $\displaystyle{\frac{\du N_{ELG}}{\du z \; \du\mathrm{deg}^2}}$ [{DESI}] & 0 & 0 & 0 & 0 & 0 & 0 & 309 & 2269 & 1923 & 2094   \\
      \cline{2-11}
       \rule{0pt}{4ex} $\displaystyle{\frac{dN_{H\alpha}}{\du z \; \du\mathrm{deg}^2}}$ [{\it Euclid}] & 0 & 0 & 0 & 0 & 0 & 0 & {2434} & {4364} & {4728} & {4825}  \\
      \botrule
    \end{tabularx}
    
    \vspace{3mm}
    \begin{tabularx}{\textwidth}{c  |  CCCCCCCCCC }
        \toprule
       $z$  & 1.05 & 1.15  & 1.25 &  1.35 & 1.45 & 1.55 & 1.65  &  1.75 & 1.85 & 1.95 \\ 
       \hline \hline
      \rule{0pt}{4ex} $\displaystyle{\frac{\du N_{LRG}}{\du z \; \du\mathrm{deg}^2}}$ [{BOSS}] & 0 & 0 & 0 & 0 & 0 & 0 & 0 & 0 & 0 & 0 \\
      \cline{2-11} 
     \rule{0pt}{4ex}
      $\displaystyle{\frac{\du N_{ELG}}{\du z \; \du\mathrm{deg}^2}}$ [{DESI}]& 1441 & 1353 & 1337 &  523 & 466 & 329 & 126 & 0 & 0 & 0 \\
      \cline{2-11}
       \rule{0pt}{4ex} $\displaystyle{\frac{dN_{H\alpha}}{\du z \; \du\mathrm{deg}^2}}$ [{\it Euclid}] & {4728} & {4507} & {4269} &  {3720} & {3104} & {2308} & {1514} & {1474} & {893} & {497}\\ 
      \botrule
    \end{tabularx}
\end{center}
    \caption{Forecasted number of target galaxies measurable by each survey: LRGs  for {BOSS}, ELGs for {DESI}, and H$\alpha$ emitters for {\it Euclid} per redshift per $\rm{deg}^2$ at each redshift bin $z$, taken from Refs.~\cite{Aghamousa:2016zmz, Amendola:2016saw, Font-Ribera2013}.} 
    \label{tab:ELGs}
\end{table*}

%%%%%%%%%%%%%%%%%%%%%%%%%%%%%%%%%%%%%%%%%%%%%%%%%%%%
\subsection{Fisher matrix}

We will obtain forecasted constraints using the Fisher-matrix formalism~\cite{Kamionkowski:1996ks,Zaldarriaga:1996xe,Tegmark:1997rp}.
For the CMB we follow the approach of Refs.~\cite{Munoz:2016owz,Galli:2014kla}. 
For the galaxy observables we detail below how we construct our Fisher matrix.

As described in Section~\ref{sec:effects}, {\limrs} suppress the clustering of matter in our universe, and as a consequence, that of biased tracers of matter, such as galaxies.
We take into account several effects to convert from matter to galaxy fluctuations.
First, there are redshift-space distortions (RSD), induced by the gravitational infall into, and peculiar velocities of galaxies~\cite{Kaiser:1987qv,Bull:2014rha}. 
We write the galaxy power spectrum as
\be
P_g(k,\mu) = \mathcal R(k,\mu) \mathcal F(k,\mu) P_{\rm cb}(k),
\ee
in terms of the power spectrum  $P_{\rm cb}(k)$ of CDM + baryon fluctuations,
where the two pre-factors $\mathcal{R}$ and $\mathcal{F}$ account for the RSD and the finger-of-god (FoG) effect, both of which make $P_g$ anisotropic, as they depend on $\mu=\hat k\cdot \hat n$, the line-of-sight angle.

We model the linear RSD term simply as
\be
\mathcal R(k,\mu) = \left [ b_1(k) + f \mu^2\right ]^2,
\ee
where $b_1$ is the linear Eulerian bias, as described above, and $f\equiv d\ln D/d\ln a$ is the logarithmic derivative of the growth factor $D$, which can be well approximated by~\cite{Linder:2005in}
\be
f(z) = \left( \dfrac{\Omega_{\rm cb} (1+z)^3}{\Omega_{\rm cb}\,(1+z)^3 +\Omega_\Lambda} \right)^\gamma,
\ee
with $\gamma=0.55$.
The non-linear FoG effect is included in the term
\be
\mathcal F(k,\mu)=\exp\left[-k^2\mu^2 \sigma_v^2/H^2\right],
\ee
with $\sigma_v = (1+z) \sqrt{ c^2 ~\sigma_z^2 + \sigma_{\rm FoG}^2/2}$,
where $\sigma_{\rm FoG} = \sigma_{\rm FoG}^{(0)} \sqrt{1+z}$, with $\sigma_{\rm FoG}^{(0)} \equiv 250 \, \rm km \, s^{-1}$ \cite{Zehavi:2001nr} as the intrinsic velocity dispersion of galaxies,
and we take a spectroscopic redshift error $\sigma_z \equiv 0.001 c$ ~\cite{Aghamousa:2016zmz}, which corresponds to the DESI precision requirement at $z=1$.

In addition, we include the Alcock-Paczynski (AP) effect \cite{Alcock:1979mp,Asgari:2016txw,Lemos:2017arq}, which accounts for changes in the observed $k$ and $\mu$ and the comoving volumes from assuming different cosmologies.
For that, we write the observed galaxy power spectrum as~\cite{MoradinezhadDizgah:2018ssw}
\be
    \tilde P_g(k',\mu') = P_g(k,\mu) \left(\dfrac{H_{\rm{true}}}{H_{\rm{fid}}} \right)
    \left(\dfrac{D_{A,\rm{fid}}}{D_{A,\rm{true}}}
    \right)^2,
    \label{eq:PgCOV}
\ee
where the subscript ``fid'' refers to fiducial, and the ``true" wavenumber $k'$ and angle $\mu'$ are given by

\begin{eqnarray}
    k'&=& k \Bigg[ (1-\mu^2) \frac{D_{A, \rm{fid}}^2(z)}{D_{A, \rm{true}}^2(z)}+\mu^2\frac{H_{\rm{true}}^2(z)}{H_{\rm{fid}}^2(z)} \Bigg]^{1/2}\\
    \mu'&=& \mu \frac{k}{k'}\frac{H_{\rm{true}}(z)}{H_{\rm{fid}}(z)}.
\label{eq:APconv}
\end{eqnarray}

Properly accounting for the AP effect, thus, implies evaluating the entire galaxy power spectrum at different wavenumbers for each cosmological-parameter change.
That can be computationally consuming, so instead we will perform a simpler step that is accurate to first order in derivatives (as any further is not captured by Fisher).
Therefore, we can write
\be
\dfrac{\partial \tilde P_g(k',\mu')}{\partial  \theta_i} = \dfrac{\partial  P_g(k,\mu)}{\partial  \theta_i} + \mathcal C_i(k),
\ee
for each parameter $\theta_i$, where 
\be
\mathcal C_i(k)= \frac{\partial P_g}{\partial k} \frac{dk}{d\theta_i} + \frac{\partial 
P_g}{\partial \mu} \frac{d\mu}{d \theta_i},
\ee
accounts for the AP correction to linear order, with the derivatives of $k$ and $\mu$ computed from Eq.~\eqref{eq:APconv}.

The Fisher element for parameters $\theta_i, \theta_j$ is then calculated as~\cite{Font-Ribera2013}
\begin{align} 
F_{ij} =& \sum_z \int k^2 \du k \int \du \mu \frac{V(z)}{2 (2\pi)^2} \left( \frac{\overline{n} \tilde P_g}{\overline{n} \tilde P_g + 1}  \right)^2  \nonumber \\
&\left( \frac{\partial \log \tilde P_g}{\partial \theta_i} \right)  \left( \frac{\partial \log \tilde P_g}{\partial \theta_j} \right),  
\end{align}
where  $V(z)$ is the comoving volume for each redshift bin summed over,
and $\overline {n}(z)$ is the comoving number density of tracers, given by
$    \overline{n}(z) =  \Delta z f_{\rm sky} V^{-1}(z) \, {dN}/({dz~d\rm deg^2}) ,$
where the last factor is reported for each survey in Table~\ref{tab:ELGs}. 
The integral over $\mu$ goes from $-1$ to $1$, and over wavenumbers from $k_{\rm min}=\pi V(z)^{-1/3}$ to $k_{\rm max}=0.2h \, \rm Mpc^{-1}$, which is mildly in the nonlinear regime~\cite{Cooray:2003dv}.
While at higher $z$ the fluctuations are smaller and, thus, we could reach higher $k_{\rm max}$ while linear, the biasing of galaxies becomes more complex, so we fix $k_{\rm max}$ for all $z$. 
We expect that non-gaussianities in the likelihood will affect constraints on cosmological parameters, but we do not model those effects in this work \cite{Hahn:2018zja}.

%%%%%%%%%%%%%%%%%%%%%%%%%%%%%%%%%%%%%%%%%%%%%%%%%%%%
\section{Results}
\label{sec:results}

In this section we discuss our cosmological constraints for a {\limr}.
We will perform two parallel analyses.
First, we will show the reach of different combinations of datasets by forecasting $\sigma(g_X)$ for a Weyl (neutrino-like) relic of different masses and temperatures, covering the entire range of interest.
Then, we will focus on the minimal case (that with $T_X^{(0)}=0.91$ K) for the four relic types we consider, and find more precisely above which mass $m_X$ they can be ruled out.

%%%%%%%%%%%%%%%%%%%%%%%%%%%%%%%%%%%%%%%%%%%%%%%%%%%%
\subsection{Full Parameter Space}

We will start with a Weyl relic, and cover a broad range of cases, where in each case we will assume that there exists a {\limr} in our universe with mass $m_X$ and temperature today $T_X^{(0)}$, and forecast how well $g_X$ can be measured as a measure of how significant a detection would be.

We  scan through a range of {\limr} masses $m_X$ from $10^{-2}$ eV, as all lighter relics behave identically, up to ${\sim} 10$ eV, where the relic abundance overcomes that of all DM. As for their temperature, we cover $T_X^{(0)}=[0.91-1.50]$ K, where the lower limit is as found in Section~\ref{sec:effects}, and the upper limit saturates the current 95\% C.L.~\Planck + {BOSS} DR12 BAO limit on $N_{\rm eff}$~\cite{Aghanim:2018eyx}.

First, as a test, we forecast the errors on $N_{\rm eff}$ by looking at our lightest relic ($m_X=0.01$ eV) as a proxy of the massless case, and translating the forecasted error $\sigma(g_X)$ in the degrees of freedom into
\be
\sigma(N_{\rm eff}) = \dfrac{\sigma(g_X)}{g_\nu} \left(\dfrac{T_X^{(0)}}{T_\nu^{(0)}}\right)^4.
\ee
For reference, we have confirmed that assuming lower values of $m_X$ result in the same forecasts for $N_{\rm eff}$.
This result is largely independent of the chosen $T_X^{(0)}$, so we will show forecasts for a Weyl fermion with $T_X^{(0)}=0.91$ K.

Beginning with the CMB, the {\it Planck}-only forecast gives $\sigma(g_X)=8.11$ corresponding to $\sigma(N_{\textrm{eff}}) = 0.19$ which is in agreement with the \Planck  value of $\sigma (N_{\rm eff})$ in non-photon radiation density when allowing extra relativistic degrees of freedom  Ref.~\cite{Aghanim:2018eyx}.
Likewise, the CMB-S4-only forecast yields $\sigma(N_{\textrm{eff}}) = 0.040$. 
This is to be compared with the value of $\sigma (N_{\rm eff})=0.035$ reported in Ref.~\cite{Abazajian:2016yjj} for the same combination of resolution and sensitivity.
The $\sim10\%$ difference is due to the delensing of T and E modes~\cite{Baumann:2015rya,Larsen:2016wpa} that is performed in Ref.~\cite{Abazajian:2016yjj} but not in our forecasts.
This is because we are chiefly interested in more massive relics, for which the phase shift is not the main cosmological signature.

In both cases, as well as the ones below, we account for a noted degeneracy with $\Sigma m_\nu$ by marginalizing over the neutrino mass in our forecasts. 
Adding LSS data only improves these results, as we show in Table~\ref{tab:Neff2}. In particular, we find that adding BOSS to {\it Planck} gives $\sigma(N_{\textrm{eff}}) = 0.14$; substituting DESI for BOSS yields $\sigma(N_{\textrm{eff}}) = 0.06$. Looking to the future, {\textrm Euclid} and CMB-S4 will lower this constraint to $\sigma(N_{\textrm{eff}}) = 0.02$.

\begin{table*}[hbtp!]
    \begin{center}
    \vspace{3mm}
        \begin{tabularx}{\textwidth}{ c | CCCC  }
        \toprule
       $\sigma ({N_{\rm eff}})$ &  CMB Only & {BOSS} & {DESI} & {\it Euclid} \\ 
       \hline \hline 
       LSS Only & \cellcolor{gray}  & 0.92 (0.84) & 0.29 (0.25) & 0.20 (0.13) \\
      %\cline{2-11}
       {\Planck} & 0.19 (0.19) & 0.14 (0.08) &  0.06 (0.04) & 0.06 (0.04) \\
      %\cline{2-11}
       {CMB-S4} & 0.04 (0.04) & 0.04 (0.03) & 0.03 (0.02) & 0.02 (0.02) \\
      \botrule
    \end{tabularx}
    \end{center}
    \caption{Forecasted 1$\sigma$ errors on $N_{\rm eff}$ from different combinations of experiments.
    Numbers in parenthesis assume fixed total neutrino mass, whereas the rest are marginalized over neutrino masses.} 
    \label{tab:Neff2}
\end{table*}

\captionsetup[figure]{justification=centerlast}
\begin{figure*}[hbtp!]
    \centering
    \includegraphics[width = 1.0\linewidth]{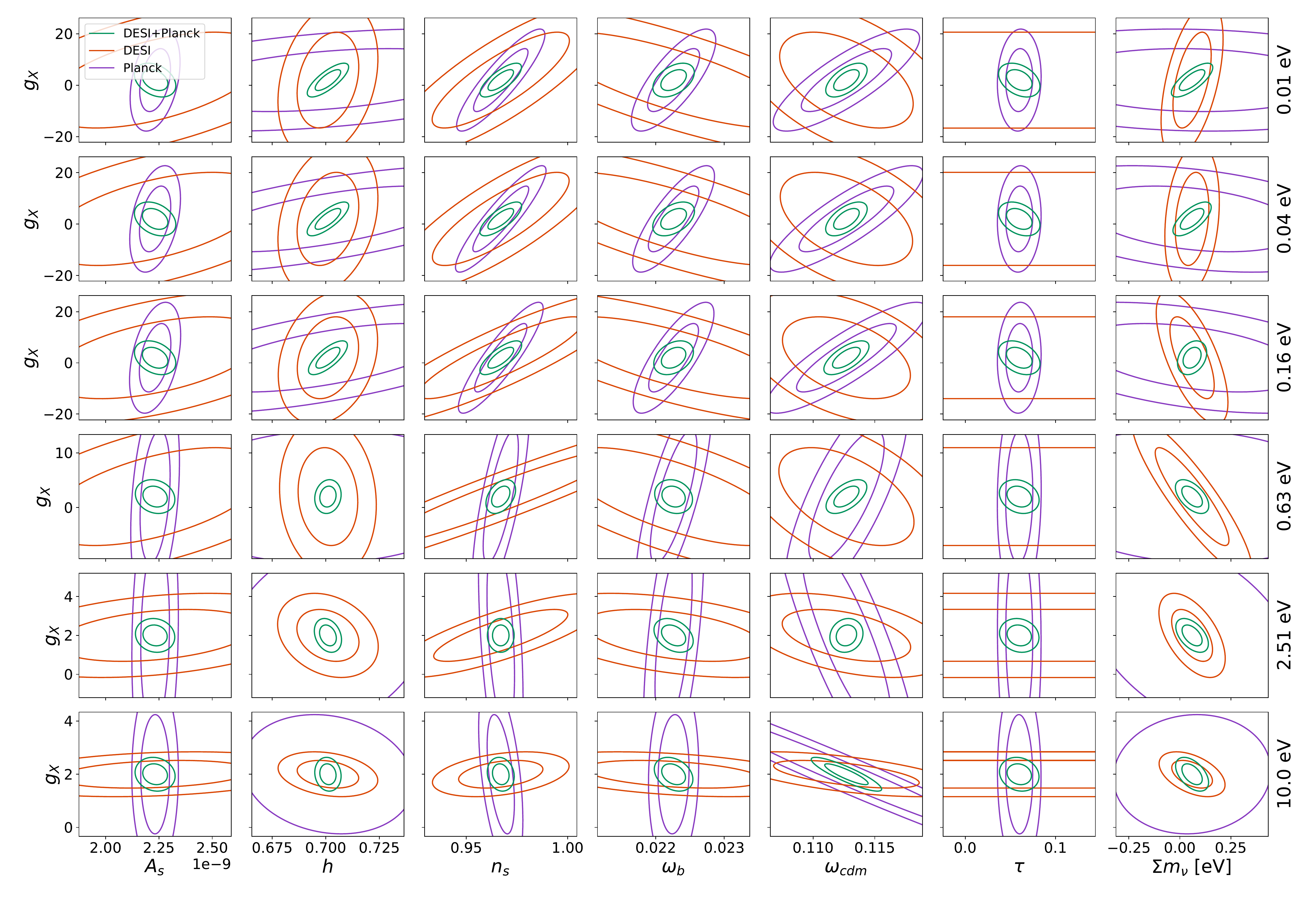}
    \caption{$68\%$ C.L. and $95\%$ C.L. projected confidence ellipses for each of the parameters we marginalize over, as well as the {\limr} number $g_X$ of degrees of freedom, for DESI (red), \Planck (purple), and their combination (green).
    Each row has a different fiducial relic mass, denoted on the right, all with an assumed temperature $T_X^{(0)}=0.91$ K at $z=0$.
    Note that we also marginalize over the unknown neutrino mass, which loosens our constraints by as much as 143\% for LSS-only information, 64\% for CMB-only information, and 81\% for combined LSS and CMB information.}
    \label{fig:squareplot}
\end{figure*}
 
We now move to non-zero masses, and provide marginalized posteriors from forecasts for a $0.91$ K (minimum temperature) Weyl relic at different masses in Fig.~\ref{fig:squareplot}.
We only show the 2D contours between $g_X$ and other cosmological parameters; for the full triangle plots at fixed mass $m_X=$ 0.01 eV, see Appendix~\ref{sec:AppB}.
The combination of information from the CMB and LSS can be seen to significantly improve constraints by breaking parameter degeneracies present in the individual datasets.
Interestingly, the degeneracy directions change with {\limr} mass.
As an example, the degeneracy line for $g_X$ and $\omega_{\rm cdm}$ for CMB data changes direction as the {\limr} becomes more massive, and starts behaving as matter instead of radiation at recombination.
The LSS degeneracy line, however, stays relatively stable, improving the CMB result by different amount at each mass.

\begin{figure}[h!]
    \centering
    \includegraphics[width = \linewidth]{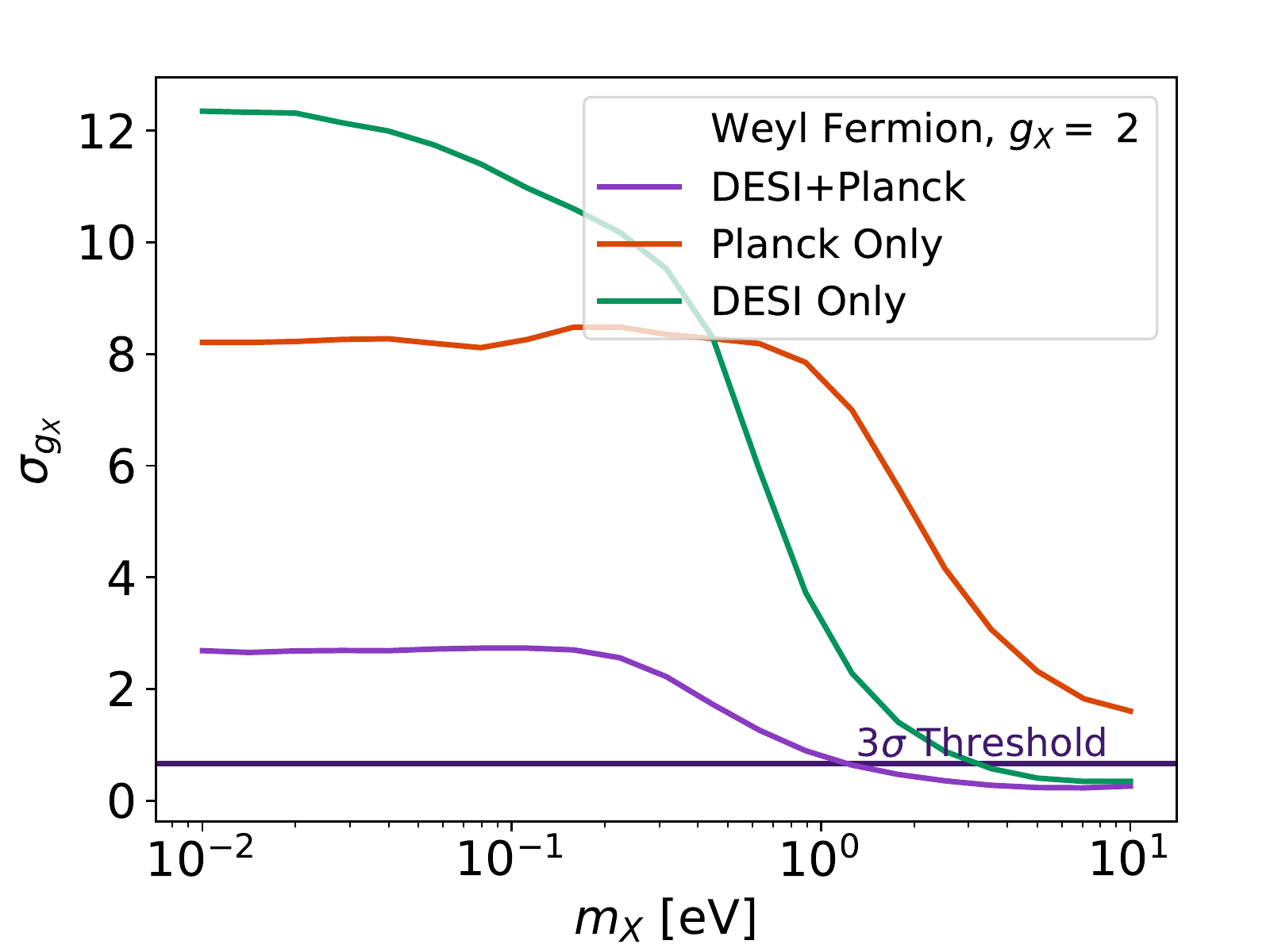}
    \caption{Improvement of Weyl relic measurements by addition of LSS data with {DESI} and \Planck constraints. The relic is fixed at its minimum possible temperature, $T_X^{(0)}=0.91$ K. As shown, the joint constraints are much stronger than the LSS or CMB alone.}
    \label{fig:LSS_improvement}
\end{figure}

The result described above indicates that combining CMB and LSS information is critical for an optimal constraint of {\limrs}.
We confirm this in Fig.~\ref{fig:LSS_improvement}, where we show the forecasted error in $g_X$ for CMB and LSS data on their own, as well as together, which dramatically improves the constraints.
For the rest of this work we will consider different combinations of CMB and LSS surveys together.

\captionsetup[figure]{justification=centerlast}

\begin{figure}[h!]
    \centering
    \includegraphics[width = \linewidth]{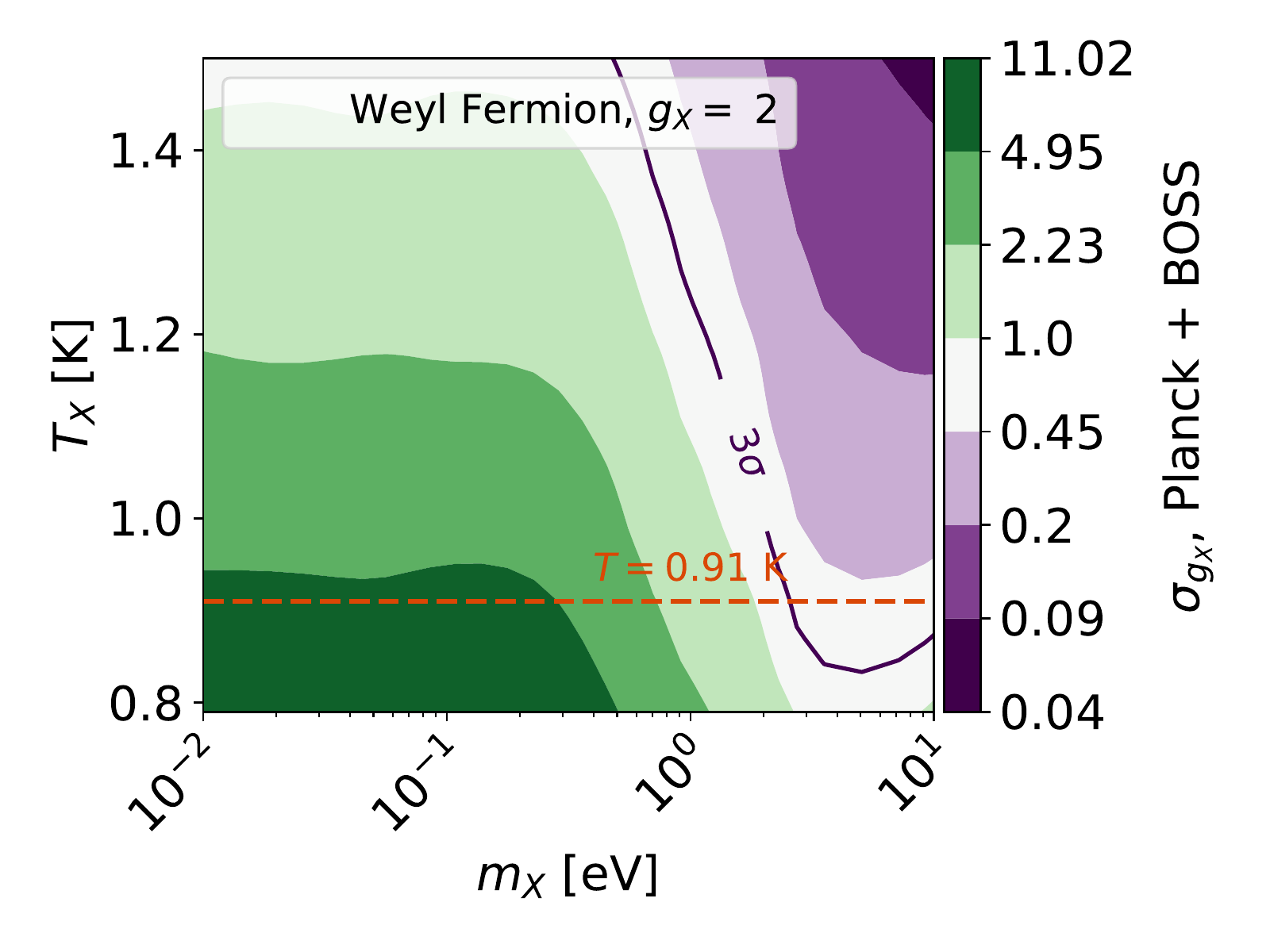}
    \caption{Forecasted errors on $g_X$ for a Weyl (neutrino-like) relic of different fiducial masses and temperatures, in all cases with fiducial $g_X = 2$, assuming {BOSS}+\Planck data. The region of parameter space measurable at the 3$\sigma$-level lays rightward of the purple solid line,
    and the dashed red line shows the minimum temperature expected for a relic.
    }
    \label{fig:Gridplot_BOSS_Planck}
\end{figure}

We now forecast to which level of significance different {\limr} can be constrained, under three different survey combinations.
The first is what would be realizable by current data, where we assume galaxy data from {BOSS} and \Planck for the CMB.
We show the forecasted $\sigma(g_X)$ in Fig.~\ref{fig:Gridplot_BOSS_Planck}, which clearly shows that {\limrs} with larger $T_X^{(0)}$ and $m_X$ are more readily observable.
However, to observe (or rule out) a {\limr} at 3$\sigma$ it has to be relatively heavy ($m_X\gtrsim$ few eV), as we will see below.
Note that in this figure we show results for $T_X^{(0)}<0.91$ K, as for instance a scalar at that minimum temperature would be equivalent to a Weyl fermion with $T_X^{(0)}=0.79$ K, as we will discuss below.

\begin{figure}[h!]
    \centering
    \includegraphics[width = \linewidth]{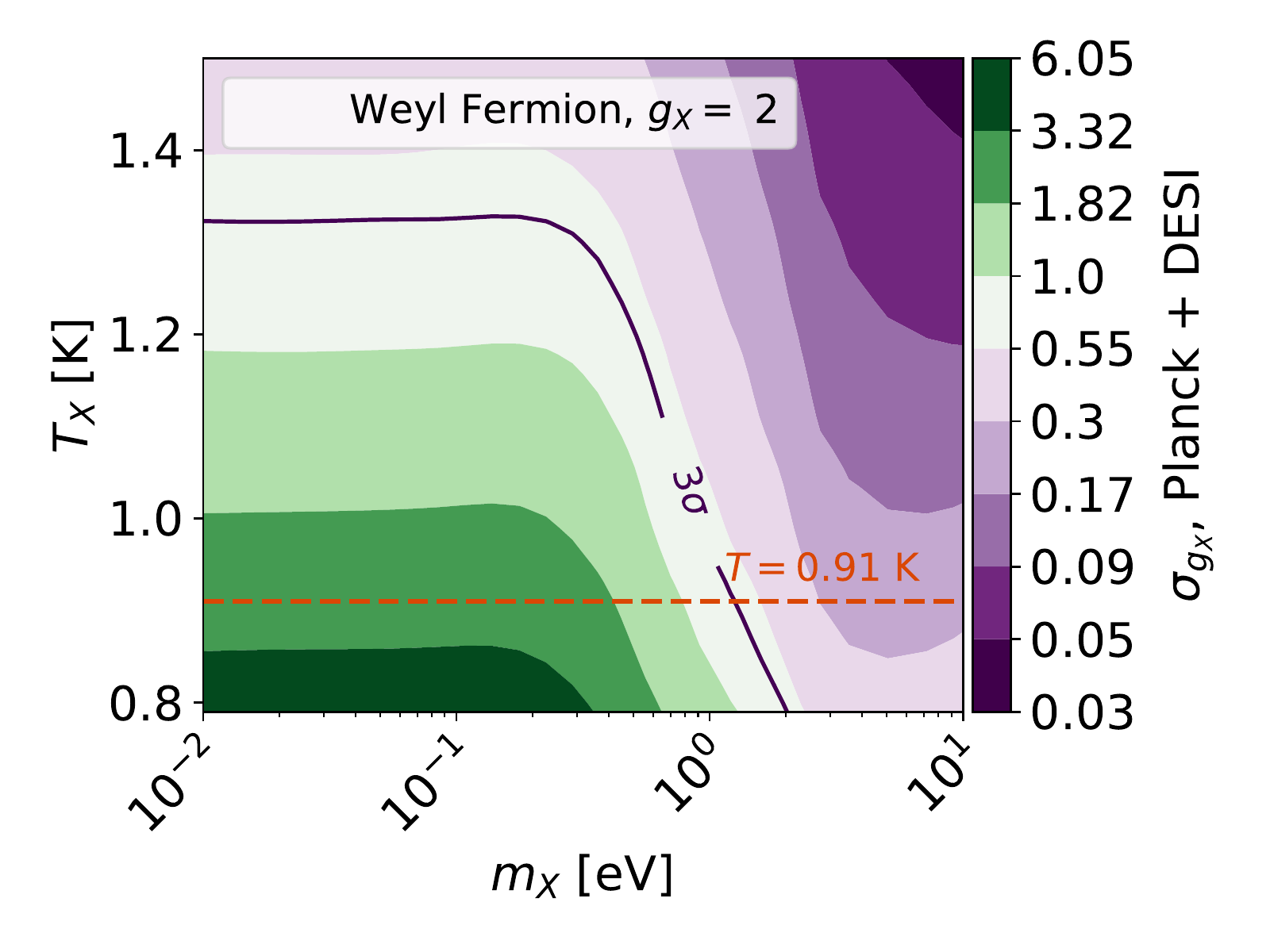}
    \caption{Same as Fig.~\ref{fig:Gridplot_BOSS_Planck} for {DESI} + \it{ Planck}.}
    \label{fig:Gridplot_DESI_Planck}
\end{figure}

The second case we consider is the near-future one, where we add {DESI} data to \Planck.
We show the forecasted constraints on $g_X$ for this combination in Fig.~\ref{fig:Gridplot_DESI_Planck}, which are clearly improved with respect to the results shown in Fig.~\ref{fig:Gridplot_BOSS_Planck}.
In this case one can rule out relics of any mass with $T_X^{(0)}=1.4$ K at 3$\sigma$.
More interestingly, we see that masses above 1 eV would be ruled out, even for the lowest possible relic temperature of $T_X^{(0)}=0.91$ K.

\begin{figure}[h!]
    \centering
    \includegraphics[width = \linewidth]{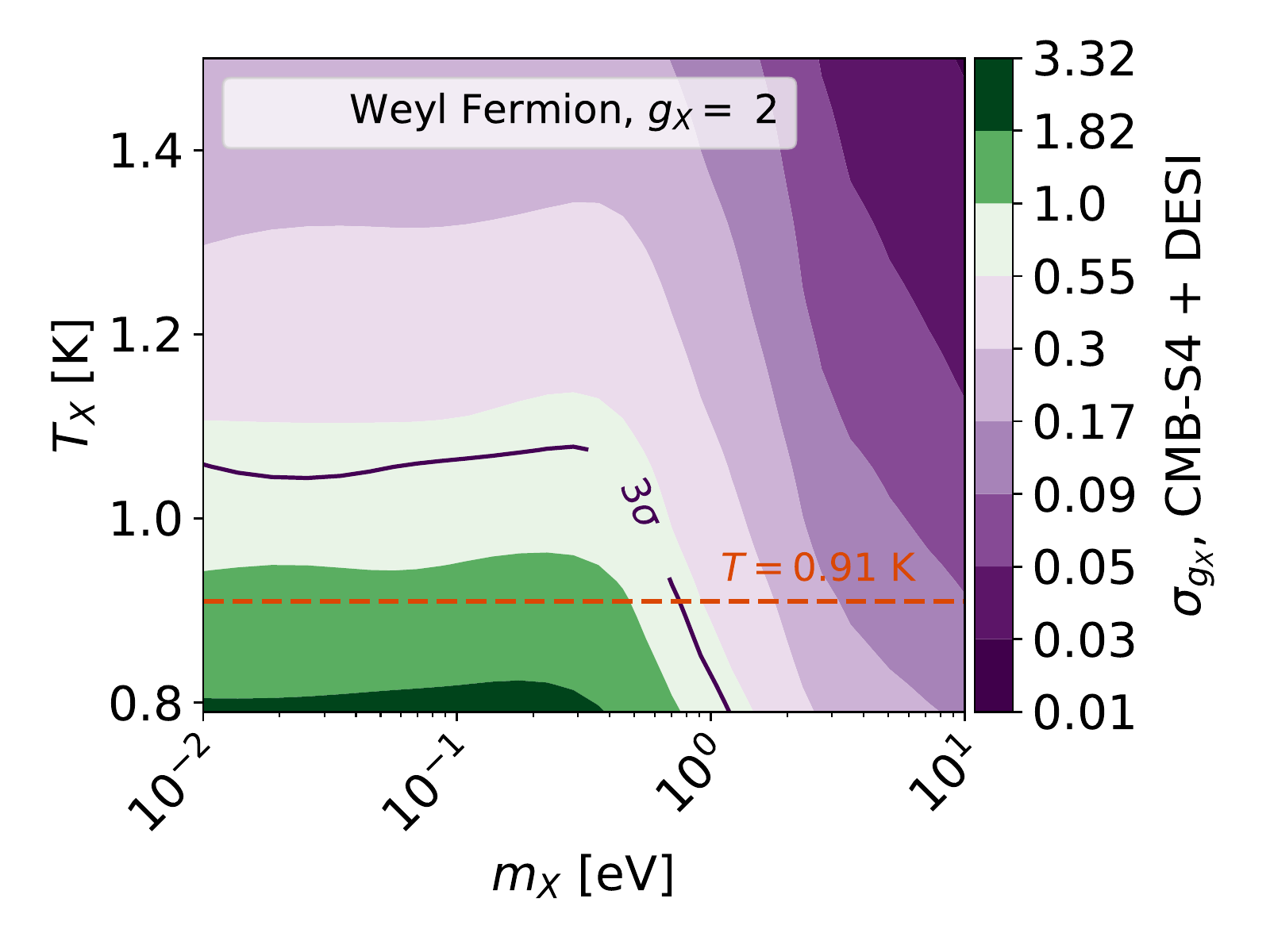}
    \caption{Same as
    Fig.~\ref{fig:Gridplot_BOSS_Planck} for {DESI} + {CMB-S4}.
    }
    \label{fig:Gridplot_DESI_CMBS4}
\end{figure}

The final case we consider is more futuristic, and adds {CMB-S4} data to {DESI}.
We show the results in Fig.~\ref{fig:Gridplot_DESI_CMBS4}, which further improves the prospects for detecting light relics.
In this case even relics at low temperatures can be ruled out at 3$\sigma$ confidence for masses above 0.78 eV, whereas minimum-temperature massless Weyl relics can only be found at 0.5$\sigma$ confidence. 

\captionsetup[figure]{justification=centerlast}

\captionsetup[figure]{justification=centerlast}

\captionsetup[figure]{justification=centerlast}

\captionsetup[figure]{justification=centerlast}
\begin{figure*}
\begin{center}
\begin{subfigure}{0.40\linewidth}
    \centering
    \includegraphics[width = \linewidth]{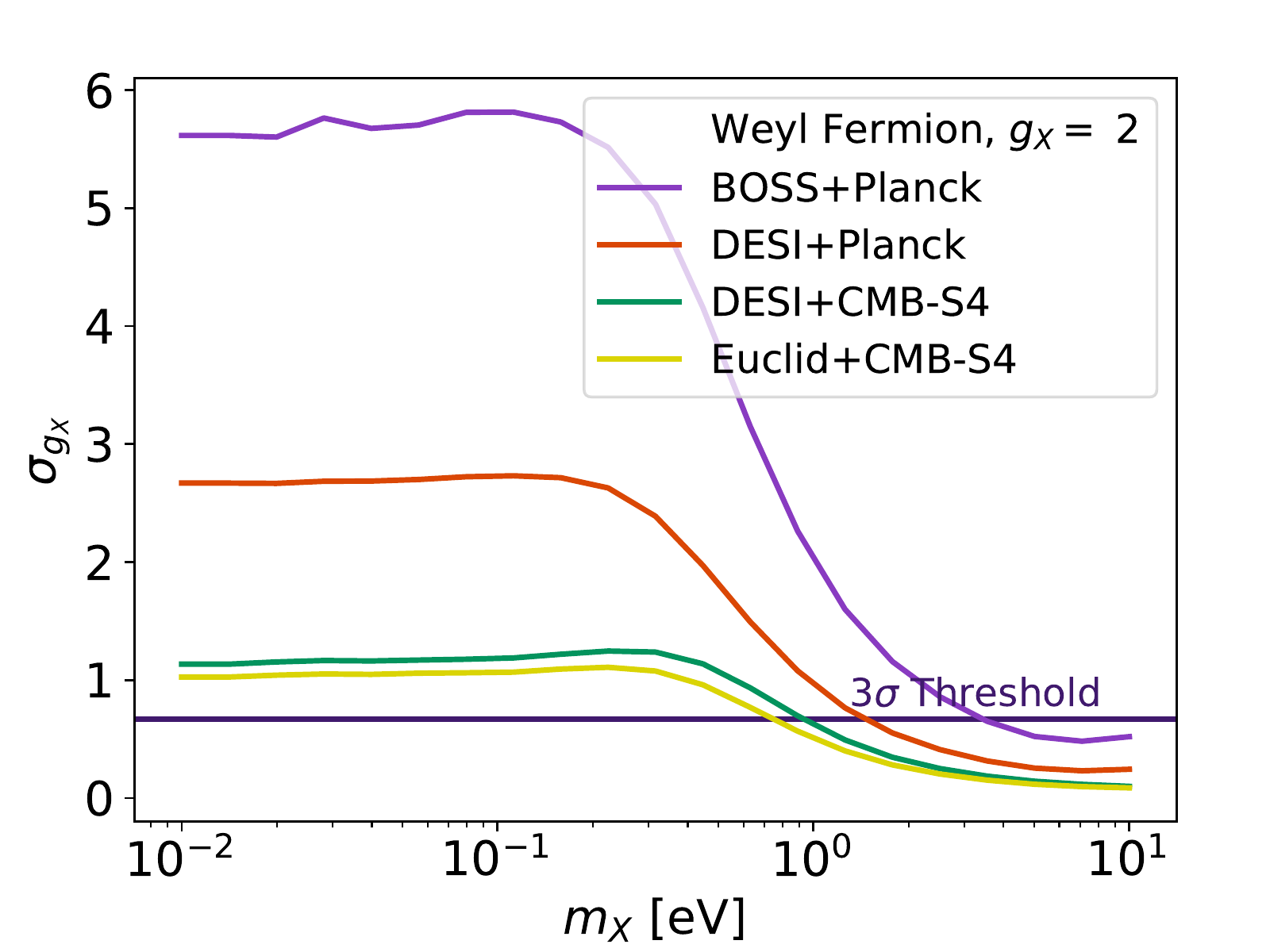}
    \caption{}
    \label{fig:Lineplot_Weyl}
\end{subfigure}
\hspace{3mm}
\begin{subfigure}{0.40\linewidth}
    \centering
    \includegraphics[width = \linewidth]{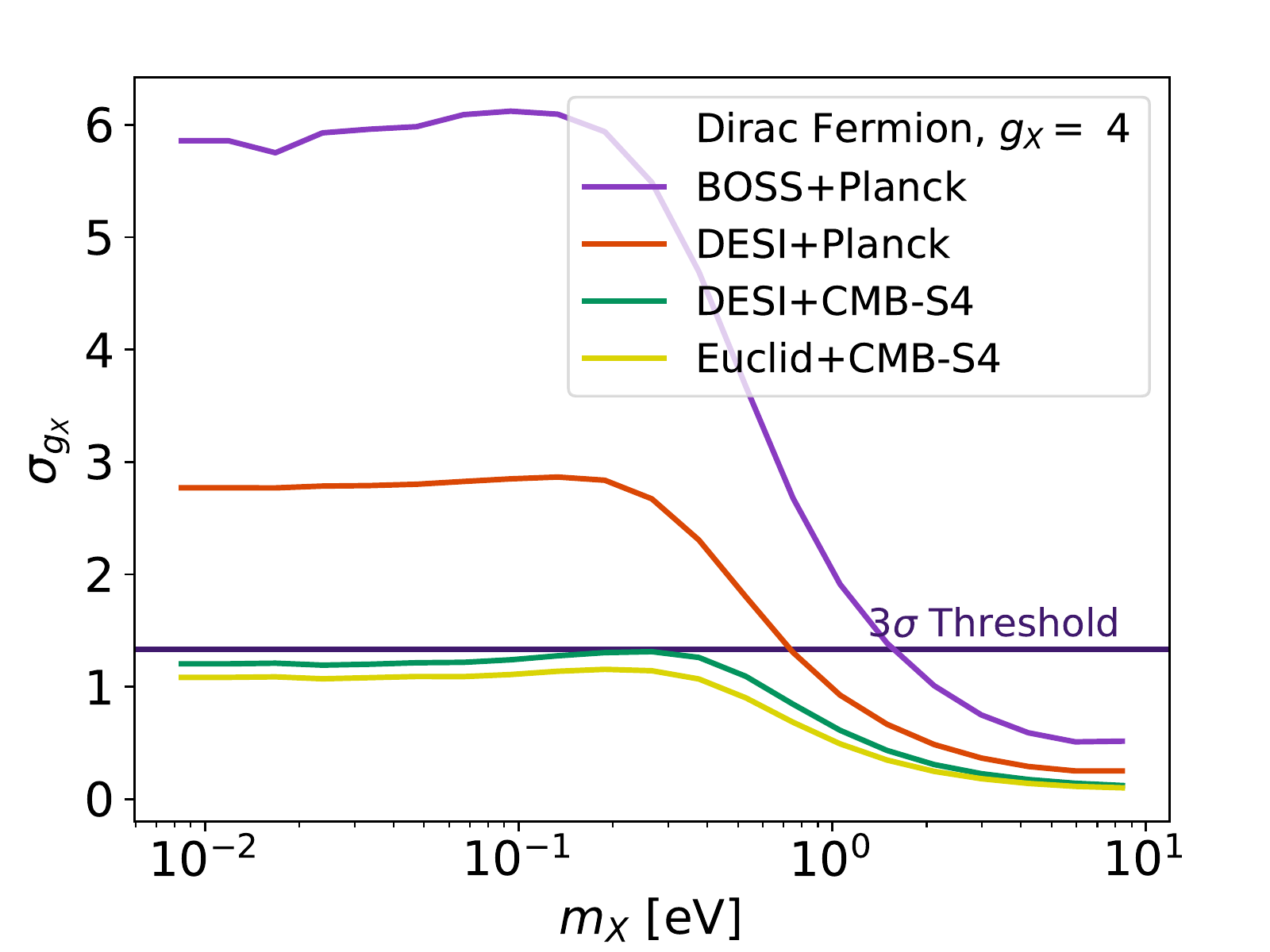}
    \caption{}
    \label{fig:Lineplot_Dirac}
\end{subfigure}
\begin{subfigure}{0.40\linewidth}
    \centering
    \includegraphics[width = \linewidth]{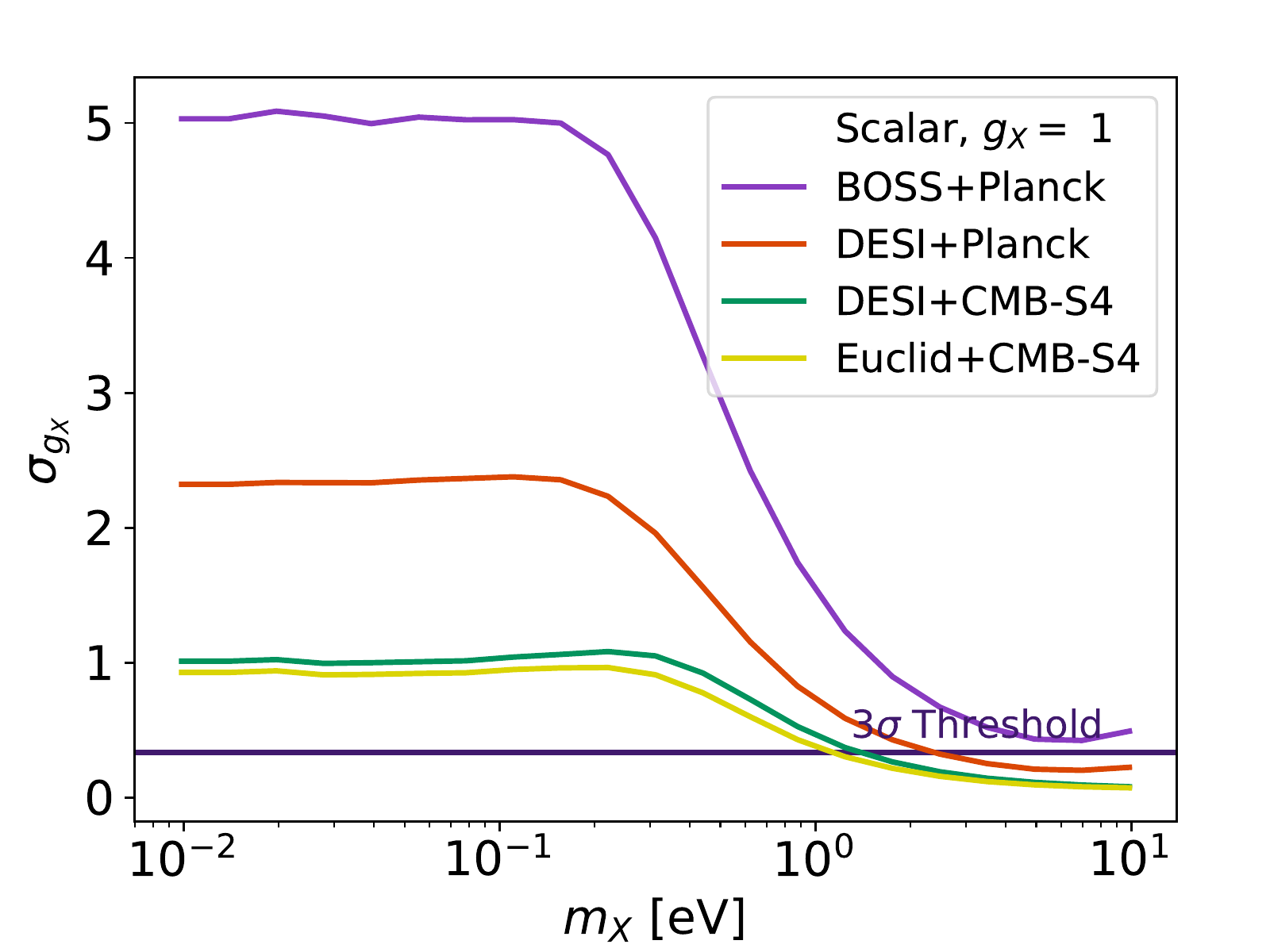}
    \caption{}
    \label{fig:Lineplot_Scalar}
\end{subfigure}
\hspace{3mm}
\begin{subfigure}{0.40\linewidth}
    \centering
    \includegraphics[width = \linewidth]{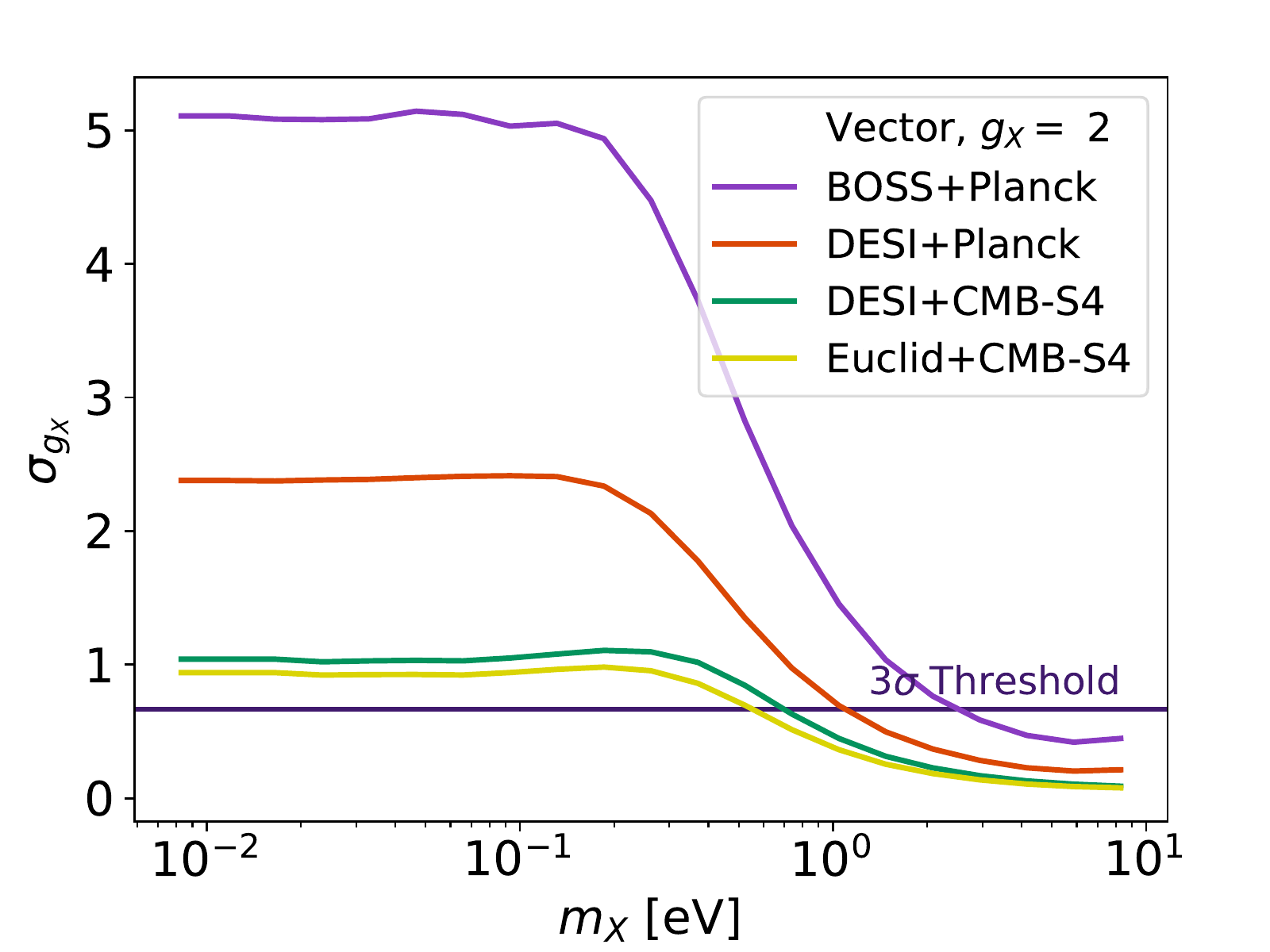}
    \caption{}
    \label{fig:Lineplot_Vector}
\end{subfigure}

\end{center}
\caption{Forecasted error on the relic degrees of freedom for a neutrino-like Weyl fermion (with fiducial $g_X = 2$, top left), a Dirac fermion ($g_X = 4$, top right),
a real scalar ($g_X = 1$, bottom left), and a vector particle ($g_X = 2$, bottom right), all at their minimum temperature $T_X= 0.91$ K, for various combinations of CMB + LSS experiments. 
The horizontal line denotes the uncertainty required to detect each relic at 3$\sigma$.
}
\label{fig:lineplots}
\end{figure*}

\captionsetup[figure]{justification=centerlast}
\begin{figure}[h!]
    \centering
    \includegraphics[width = \linewidth]{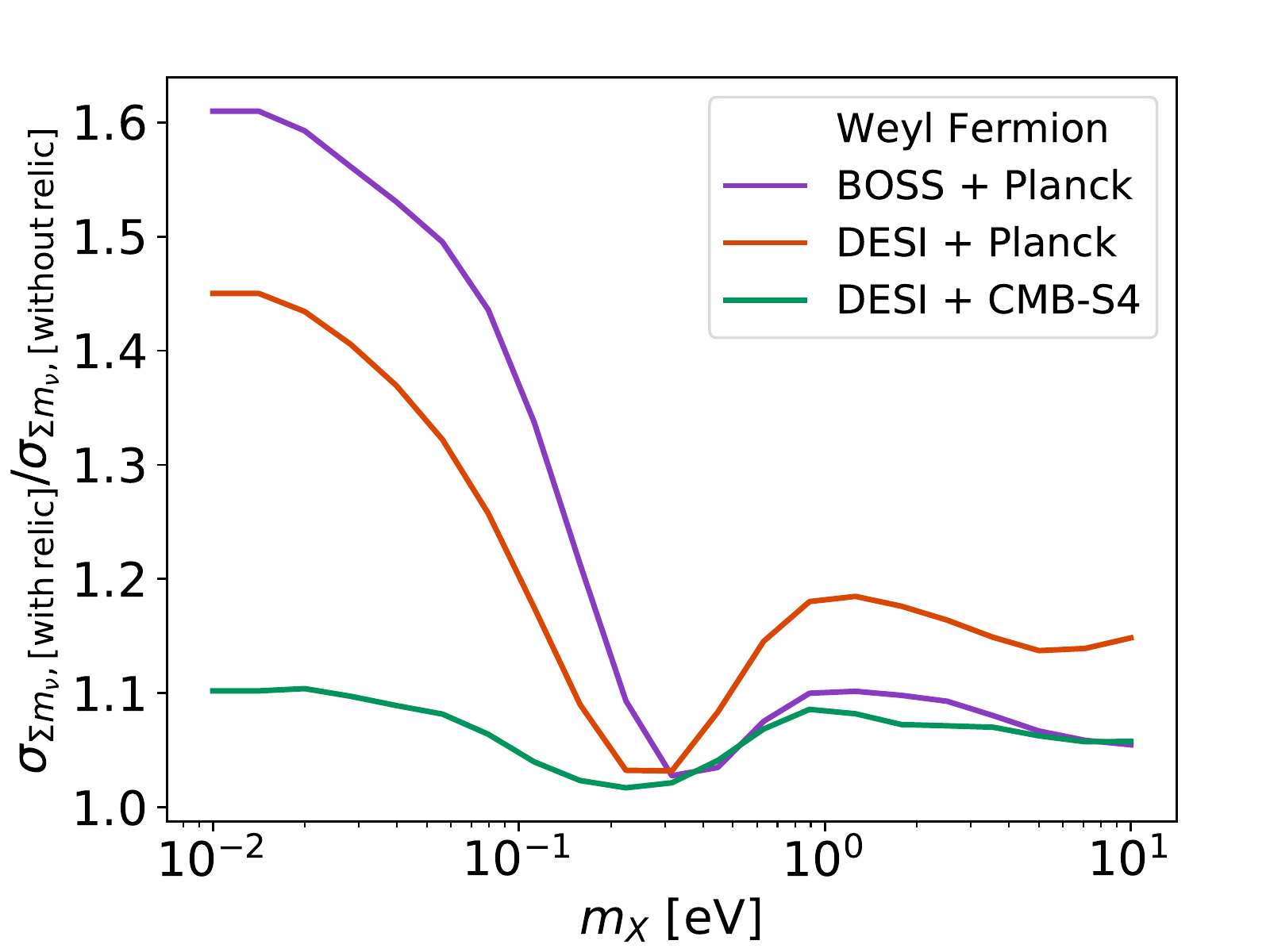}
    \caption{Forecasted {DESI} + {CMB-S4} uncertainty on the sum $\sum m_\nu$ of neutrino masses, when it is jointly searched for with a relic of mass $m_X$ and temperature $T_X=0.91$ K. The degeneracy is minimized at $\sim \mathcal{O}(0.3 {\rm eV})$ for all particle types, although the constraints on neutrino masses using CMB  data from \Planck are always expected to weaken by $\sim 10\%$, if a new light relic is present.
    }
    \label{fig:Mnu_Sigmas}
\end{figure}

%%%%%%%%%%%%%%%%%%%%%%%%%%%%%%%%%%%%%%%%%%%%%%%%%%%%
\subsection{Minimum Temperature}

While the figures discussed above covered a broad range of temperatures and masses, they all assumed a Weyl relic. Here we extend our results to other types of relics, focusing on the minimum temperature of $T_X^{(0)}=0.91$ K,
corresponding to the earliest decoupling from the SM plasma. 
We divide our results into fermionic and bosonic relics.
The cumulative results of our forecast for each type of particle are tabulated in Table~\ref{tab:constraints}.

%%%%%%%%%%%%%%%%%%%%%%%%%%%%%%%%%%%%%%%%%%%%%%%%%%%%
\subsubsection*{Fermionic Relics}

We start with a massive Weyl fermion with $T_X^{(0)}=0.91$ K, for which we show our forecasts on $\sigma({g_X})$ for various combinations of galaxy surveys and CMB experiments in Fig.~\ref{fig:Lineplot_Weyl}, with a finer mass resolution than the results above. 
We report the minimum relic masses that are
observable at 3$\sigma$ significance, both with (and without) marginalizing over the neutrino masses, as a test of how degenerate {\limrs} are with the total neutrino mass. 
The combination of presently available \Planck and {BOSS} datasets are forecasted to observe or rule out {\limrs} above 2.85 (2.47) eV at 3$\sigma$ significance. For \Planck and {DESI}, this is lowered to {\limrs} with masses above 1.20 (1.00) eV. This result should motivate an analysis using presently available datasets. For the futuristic combination of {CMB-S4} and {\it Euclid} datasets, we show that {\limr} masses above 0.63 (0.59) eV can be observed or ruled out at 3$\sigma$ significance.

As an example of the physical implications of these constraints, let us apply to them to the ($s=3/2$) gravitino, which is related to the scale of SUSY breaking in some models.
The gravitino is cosmologically equivalent to the neutrino-like Weyl relic that we have studied, as only the $s=1/2$ modes are thermalized with the SM plasma in the early universe~\cite{Osato:2016ixc}, and are expected to have the lowest relic temperature of 0.91 K.
This has allowed previous work to constrain the gravitino mass by requiring that their abundance does not overcome that of the cosmological dark matter~\cite{Moroi:1993mb}.
Our forecast above shows that current data is sensitive to gravitinos heavier than $m_X=2.85$ eV, which is around the benchmark of some models of SUSY breaking~\cite{Hook:2015tra,Hook:2018sai}, 
and a factor of a few better than the best limits currently available~\cite{Viel:2005qj,Osato:2016ixc}.
Upcoming data from CMB-S4 combined with {\it Euclid} is expected to further detect such gravitino population masses above 0.63 eV. 
Under the assumption that a cosmological gravitino population no longer exchanges entropy after decoupling from the SM bath, we can relate constraints on $m_X$ to bounds on the SUSY breaking scale $\Lambda_{\rm SUSY} \sim \sqrt{m_X M_{\rm Pl}}$~\cite{Deser:1977uq,Martin:1997ns}.
Our forecasted {\it Planck} and BOSS dataset translates to an upper bound $\Lambda_{\rm SUSY} \lesssim 80$ TeV, whereas the CMB-S4 and {\it Euclid} datasets lower this to $\Lambda_{\rm SUSY} \lesssim 50$ TeV. 
These projections are interestingly complementary to the energy range that will be reached by the proposed $\mathcal{O}(100 \textrm{ TeV})$ particle collider,
showing the promise of our approach.
 
We also consider a Dirac fermion, with $g_X=4$ and mass $m_X$.
In terms of the equivalent Weyl fermion, this corresponds to a temperature $T_W^{\rm eq} = 1.08$ K and mass $m_W^{\rm eq} = 1.19\, m_X$. In Fig.~\ref{fig:Lineplot_Dirac}, we show that the combined \Planck and {BOSS} datasets are forecasted to observe or rule out such particles above 1.30 (1.12) eV at 3$\sigma$ significance. For \Planck and {DESI}, the 3$\sigma$ constraint is lowered to 0.61 (0.52) eV. Interestingly, CMB-S4 data will enable the  parameter space of Dirac fermions with any mass to be observed or ruled out at 3$\sigma$ significance when combined with LSS data from DESI. 

%%%%%%%%%%%%%%%%%%%%%%%%%%%%%%%%%%%%%%%%%%%%%%%%%%%%
\subsubsection*{Bosonic Relics}

We now move to bosonic degrees of freedom.
First, we study a minimum-temperature real scalar, with $s=0$, $g_X=1$, and mass $m_X$.
This is equivalent to a Weyl relic with $T_W^{\rm eq} = 0.79$ K and $m_W^{\rm eq} = 1.01 m_X$.
We show in Fig.~\ref{fig:Lineplot_Scalar} that, while the combination of presently available \Planck and {BOSS} datasets cannot constrain scalar relics at the 3$\sigma$ significance, {DESI} and \Planck can jointly rule out scalars with masses above 1.96 (1.61) eV.  
Further,  the combination of {CMB-S4} with either the {DESI} or {\it Euclid} datasets can observe or rule-out real scalar bosonic relics above 1.14 (1.06) and 0.93 (0.87)  eV, respectively.

Second, we consider a massive vector, with $s=1$ and $g_X=2$.
This massive vector is equivalent to a Weyl relic with $T_W^{\rm eq} = 0.94$ K and $m_W^{\rm eq} = 1.21 m_X$.
In Fig.~\ref{fig:Lineplot_Vector} we show that the combination of \Planck and {BOSS} datasets can observe or rule-out massive vector bosonic relics above 2.05 (1.79) eV, whereas substituting BOSS for DESI improves this number to 0.90 (0.75) eV. Combining the {CMB-S4} and {\it Euclid} datasets further improves this to 0.47 (0.44) eV.

\begin{table*}[hbtp!]
    \begin{center}
    \vspace{3mm}
        \begin{tabularx}{\textwidth}{ c | CCCC  }
        \toprule
         \multicolumn{1}{c}{}   &  CMB Only & {BOSS} & {DESI} & {\it Euclid} \\ 
       \hline \hline
          \multicolumn{1}{c}{} & \multicolumn{4}{c}{Scalar $m_X$[eV]}\\
         \hline
       LSS Only &  \cellcolor{gray} & - (-) & 4.98 (4.54) & 3.24 (3.22) \\
       {\it Planck} &  - (-) & - (-) & 1.96 (1.61) & 1.31 (1.16)   \\
       {CMB-S4} &  1.48 (1.44) & 1.41 (1.31) & 1.14 (1.06) & 0.93 (0.87)   \\
       \hline \hline
          \multicolumn{1}{c}{} & \multicolumn{4}{c}{Weyl Fermion $m_X$[eV]}\\
         \hline
       LSS Only & \cellcolor{gray}  & - (-) & 3.13 (2.78) & 2.42 (2.41) \\
       {\it Planck} &  - (-) & 2.85 (2.47) & 1.20 (1.00) & 0.87 (0.78)   \\
       {CMB-S4} &  1.03 (1.02) & 0.98 (0.91) & 0.78 (0.71) & 0.63 (0.59)   \\
       \hline \hline
          \multicolumn{1}{c}{} & \multicolumn{4}{c}{Vector $m_X$[eV]}\\
         \hline
       LSS Only &  \cellcolor{gray} & - (-) & 2.41 (2.08) & 1.88 (1.88) \\
       {\it Planck} &  - (-) & 2.05 (1.79) & 0.90 (0.75) & 0.65 (0.60)   \\
       {CMB-S4} &  0.81 (0.78) & 0.75 (0.70) & 0.58 (0.54) & 0.47 (0.44)   \\
       \hline \hline
          \multicolumn{1}{c}{} & \multicolumn{4}{c}{Dirac Fermion $m_X$[eV]}\\
         \hline
       LSS Only & \cellcolor{gray}  & 4.06 (3.72) &  1.82 (1.36) &  1.50 (1.50) \\
       {\it Planck} &  - (-) & 1.30 (1.12) & 0.61 (0.52) & 0.45 (0.43)   \\
       {CMB-S4} &  0.56 (0.55) & 0.51 (0.48) & \textbf{All} (\textbf{All}) & \textbf{All} (\textbf{All})   \\
      \botrule
    \end{tabularx}
    \end{center}
    \caption{Minimum mass at which a {\limr} (scalar boson, Weyl fermion, vector boson or Dirac fermion, from top to bottom) can be observed or ruled out at 3$\sigma$ significance. Also reported in parentheses is the result with fixed $\sum m_\nu$ (to its fiducial value). A``$-$'' sign corresponds to no masses within the 3$\sigma$ constraint. ``\textbf{All}'' corresponds to all {\limr} masses analyzed being within the 3$\sigma$ constraint.  
    } 
    \label{tab:constraints}
\end{table*}

%%%%%%%%%%%%%%%%%%%%%%%%%%%%%%%%%%%%%%%%%%%%%%%%%%%%
\subsection{Neutrino-mass forecasts}

We have detailed in each previous subsection the constraints with and without marginalizing over neutrino masses to emphasize the importance of this step, as it is seen to affect results noticeably when LSS information is being considered. We note that DESI is particularly sensitive to the marginalization or fixing of $\sum m_\nu$. This is due to its chosen bias prescription, which does not include a parameter to marginalize over the redshift dependence of the bias, as opposed to BOSS and \textit{Euclid}. 
This underscores the sensitivity of our results to the details of the bias prescription, which is further explored in our companion paper~\cite{Xu:2020fyg}.

As a consequence of our analysis, we can also forecast how much neutrino-mass measurements would be affected by the presence of a {\limr}, given the degeneracies between
$\sum m_\nu$ and $g_{X}$ shown in Fig.~\ref{fig:squareplot}. 
We show in Fig.~\ref{fig:Mnu_Sigmas} the relative increment in the error of the sum $\sum m_\nu$ of neutrino masses when marginalizing over a relic of varying mass. For reference, we forecast $\sigma( \sum m_\nu)$ to be $61.1 \times 10^{-3}$ eV for BOSS and {\it Planck}, $28.2 \times 10^{-3}$ eV for DESI and {\it Planck}, and $24.1 \times 10^{-3}$ eV for DESI and CMB-S4, with a fiducial at the (normal-hierarchy) minimum $\sum m_\nu = 60 \times 10^{-3}$ eV and no other relics.  
The degradation in the expected errors ranges from 10\% for heavy relics and futuristic data (DESI+S4), to nearly 100\% for lower masses and current or upcoming data.   (BOSS/DESI+{\it Planck}). 
Note that for relics of $m_X\approx 0.3$ eV the degradation minimizes in all survey specifications.
This mass corresponds to relics that become non-relativistic around the time of recombination.
In essence, heavier relics produce suppression in the matter fluctuations, whereas lighter relics chiefly affect CMB and LSS observables through their change in $N_{\rm eff}$.
We encourage the reader to see our companion paper~\cite{Xu:2020fyg} for in-depth neutrino forecasts without relics.

%%%%%%%%%%%%%%%%%%%%%%%%%%%%%%%%%%%%%%%%%%%%%%%%%%%%
\section{Conclusions}
\label{sec:conclusions}

In this work we have studied how well current and upcoming cosmological surveys can detect light (but massive) relics ({\limrs}), focusing on the $10^{-2}$ eV to $10^1$ eV mass range.
These particles become non-relativistic before $z=0$, and thus affect the formation of structures in the universe.
By combining information from the CMB and the LSS we have shown that a large swath of the 2D-parameter space (of relic mass and temperature) will be probed by upcoming surveys. 

There is a minimum temperature that any relic that was in thermal equilibrium with the Standard Model should have, $T_X^{(0)}=0.91$ K.
Interestingly, we find that Weyl, vectors, and Dirac relics with this temperature, and masses above $\approx 1$ eV, can be observed or ruled out at the 3$\sigma$ significance using the presently available combination of {\it Planck} and {BOSS} datasets. 
Looking slightly to the future, the {\it Planck} and {DESI} datasets will improve these constraints, and reduce the minimum mass allowed for {\limrs} by roughly $50\%$. 
The more futuristic {\it Euclid} and {CMB-S4} datasets will present an 80\% improvement and, in the case of Dirac fermions, fully cover the parameter space. If the sum of neutrino masses, $\sum m_\nu$, can be learned independently of CMB and LSS surveys, the effect of fixing the $\sum m_\nu$ parameter manifests as an approximate 20\% improvement on these constraints. This could be accomplished, for example, by KATRIN which currently sets the leading upper bound on the effective electron neutrino mass of $1.1$ eV, independently of cosmology \cite{Aker:2019uuj}. We emphasize that the effect of marginalizing $\sum m_\nu$ significantly weakens the 3$\sigma$ constraints for some of the cases reported, suggesting that it is important to account for $\sum m_\nu$ in any search for {\limrs}. While the need to properly account for $\sum m_\nu$ has been discussed in previous work~\cite{Baumann:2018qnt, Green:2019glg, Baumann:2017gkg, Dodelson:2016wal, Font-Ribera:2013rwa, Bashinsky:2003tk}, our analysis, which does so for massive but light relics, is unprecedented. 

This result is particularly interesting for the case of the gravitino. Since the gravitino would have a cosmological imprint identical to a Weyl fermion, we have shown that \Planck and BOSS can observe or rule out gravitinos heavier than 2.85 eV. If a gravitino, or any other {\limr}, were detected, then their parameters (i.e., mass and temperature) could also be measured, as suggested in  Ref.~\cite{Banerjee:2016suz}.

In summary, while light relics are commonly assumed to be nearly massless --- and constrained through $N_{\rm eff}$ --- here we have shown that relics with masses on the $10^{-2}$ eV to $10^1$ eV scale can be constrained with cosmological data. These constraints are broadly expected to apply to the full range of allowed relic masses, from effectively massless to saturating  the DM abundance.  This complements current efforts in the search of relics, allowing many new routes for finding physics beyond the Standard Model.

%%%%%%%%%%%%%%%%%%%%%%%%%%%%%%%%%
\section*{Acknowledgments}

We thank Sunny Vagnozzi for insightful comments on a previous version of this manuscript.
We also thank Prateek Agrawal and David Pinner for discussions.
ND was supported by a National Physical Science Consortium Graduate Fellowship for STEM Diversity.
CD and JBM were partially supported by NSF grant AST-1813694.

%%%%%%%%%%%%%%%%%%%%%%%%%%%%%%%%%
\bibliography{relics_forecasts}

%%%%%%%%%%%%%%%%%%%%%%%%%%%%%%%%%
\appendix
\section{MCMC Validation of Fisher Forecasts}
\label{sec:AppMCMC}

\begin{figure*}
    \centering
    \includegraphics[width = 1.0\linewidth]{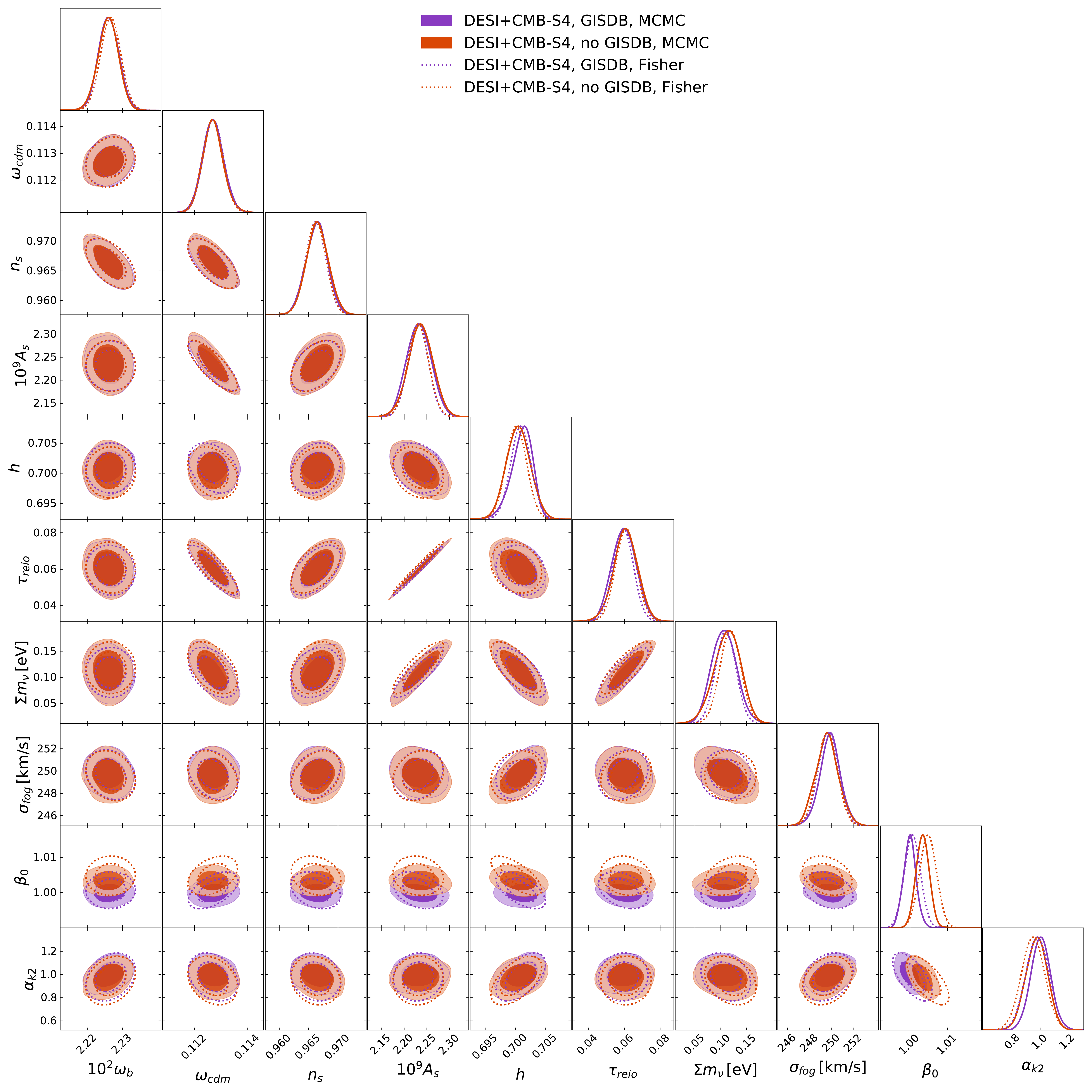}
    \caption{MCMC and Fisher forecasted marginalized posteriors for cosmological parameters and nuisance parameters for a joint {DESI} + {CMB-S4} analysis. The degenerate hierarchy is assumed with a total mass of $\sum m_\nu = 0.1$ eV. Models with and without the bias step (GISDB) are considered. As shown, the good consistency between MCMC and Fisher results, particularly the reproduced shift in parameters upon turning off GISDB, demonstrates that the effects we consider are well-captured at linear order and validates our results regarding the detectability of {\limrs}.
    } 
    \label{fig:convergence}
\end{figure*}

In this Appendix we show a comparison of our Fisher formalism and an MCMC analysis of the same mock data to confirm our Fisher analysis throughout the main text. 
In Fig.~\ref{fig:convergence} we show the MCMC (solid) and Fisher forecasted (dotted) marginalized posteriors for cosmological parameters and nuisance parameters (including the neutrino mass $\sum m_\nu$), assuming {CMB-S4} + {DESI} data.
This Figure shows that the predicted errors agree remarkably well between our Fisher-matrix approach and the full MCMC of mock data.

\captionsetup[figure]{justification=centerlast}

\captionsetup[figure]{justification=centerlast}
\begin{figure*}[]
    \centering
    \includegraphics[width = 1.0\linewidth]{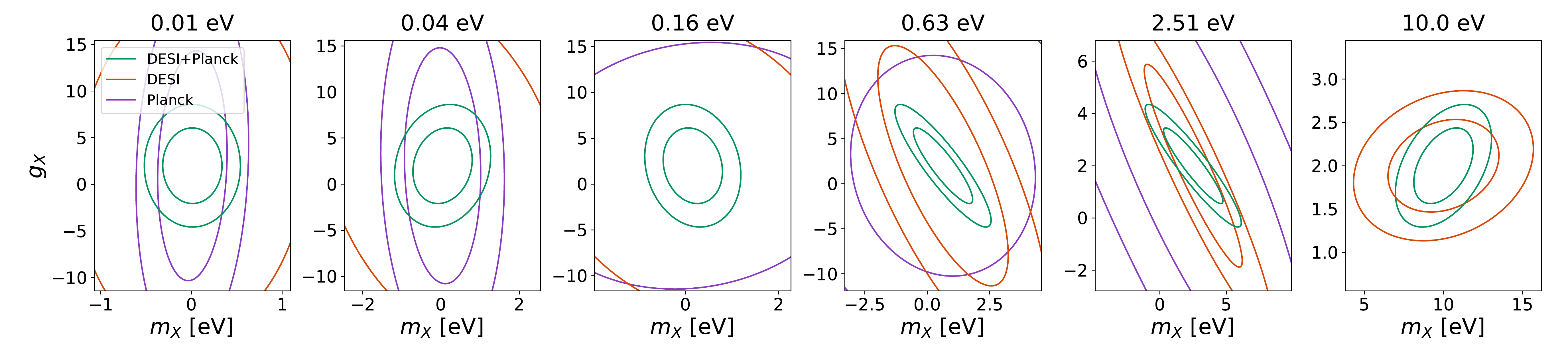}
    \caption{Fisher-matrix forecasted marginalized posteriors for the parameters $g_X$ and $m_X$. In this forecast, the LiMR mass has been allowed to vary in addition to its degrees of freedom. We present the marginalized posterior contours for five choices of the fiducial LiMR mass: $10^{-2}$ eV, $10^{-1.4}$ eV, $10^{-0.8}$ eV, $10^{-0.2}$ eV,  $10^{0.4}$ eV and $10^{1}$ eV. As shown, the degeneracy lines are driven by the relative orthogonality of CMB information at low masses, and by strong degeneracy in the LSS data at intermediate to high masses.
    }
    \label{fig:MchiMarg2}
\end{figure*}

\captionsetup[figure]{justification=centerlast}
\begin{figure}[]
    \centering
    \includegraphics[width = 1.0\linewidth]{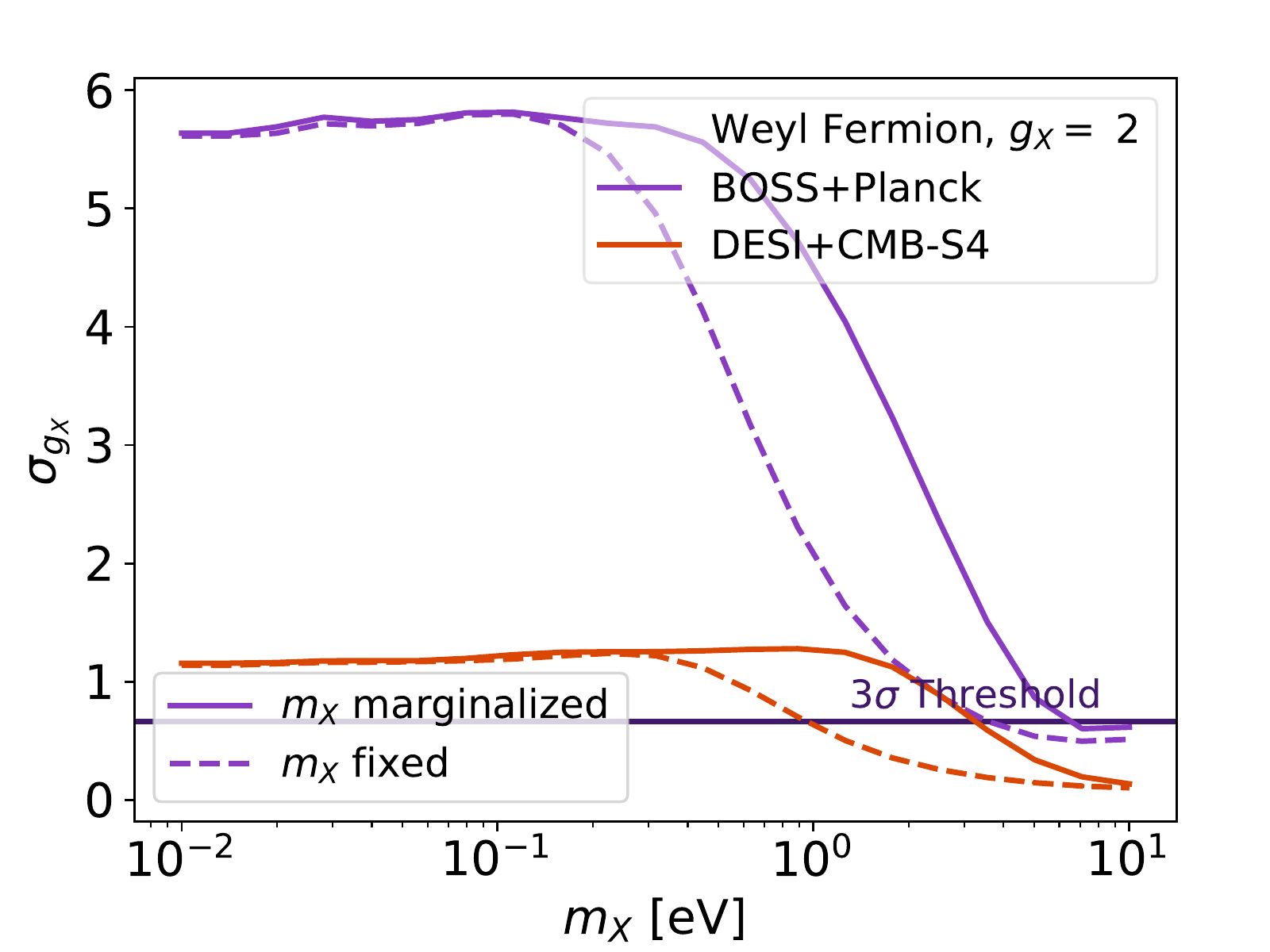}
    \caption{Forecasted sensitivity on the relic degrees of freedom $g_X$ for a Weyl fermion with (solid) and without (dashed) marginalization over relic mass $m_X$ for combinations of datasets BOSS + \Planck and DESI + CMB-S4. As expected, there is little degeneracy in the low-mass limit, where the relic mainly contributes as $N_{\rm eff}$; the $g_X-m_X$ degeneracy enters most at intermediate masses.}
    \label{fig:MchiMarg1}
\end{figure}

Moreover, we show posteriors for models with and without the growth induced scale-dependent modification to the bias (as described in our companion paper~\cite{Xu:2020fyg}), which we termed GISDB. 
The MCMC results are from Ref.~\cite{Xu:2020fyg}, and the Fishers are calculated here.
The non-GISDB Fisher ellipses are centered on the corresponding MCMC maximum likelihood point. 
The GISDB ones, however, are shifted by~\cite{Munoz:2015fdv}
\be 
\delta \theta_i = (F^{-1})_{ij} D_j,
\ee
in each parameter $\theta_i$, where we have defined
\begin{align} D_{j} &= \sum_z \int k^2 \du k \int \du \mu \frac{V(z)}{2 (2\pi)^2} \left( \frac{\partial \log \tilde P_g(k, \mu)}{\partial \theta_j} \right)  \nonumber \\
&\left( \tilde P_{g, \text{GISDB}}(k,\mu) - \tilde P_{g, \text{no GISDB}}(k,\mu)  \right)  \left( \frac{\overline{n} \tilde P_g}{\overline{n} \tilde P_g + 1}  \right)^2 ,  \end{align}
and the GISDB Fisher ellipses are computed centered on the shifted best-fit. As shown, the good cohesion between the Fisher and MCMC analyses of the data, particularly in the inclusion of the GISDB effect, demonstrates that the considered effects are well-approximated by the linearity of the Fisher approach, and thus validates the constraints we present on additional light relics.

\section{Marginalization Over the Relic Mass}
\label{sec:AppMarg}

Throughout the main text, the LiMR mass has been held fixed. In this appendix, we allow the LiMR mass to vary in the forecasts to study what effect this has on the LiMR constraints presented earlier, as well as to study how well a prospective LiMR detection could constrain its properties. 

For all combinations of LiMR species, galaxy surveys and CMB experiments studied in this work, we find that marginalizing over the LiMR mass $m_X$ weakens the constraint on the relic degrees of freedom $g_X$, as expected. 
This effect is most exaggerated in the cases where the constraint is dominantly set by LSS information.
In a joint {\it Planck}-BOSS analysis,  high-mass relics (with $m_X \geq 0.2$ eV) see the $g_X$ constraint weakened by nearly a factor of 2. 
In cases where CMB information dominates, however, such as when adding CMB-S4 to BOSS,  the $g_X$ constraint is weakened by no more than 6\%. 
Adding Planck information to DESI, the higher-mass region sees the $g_X$ constraint weakened by no more than a factor of 2. Adding CMB-S4 to DESI, the $g_X$ constraint is weakened by no more than 25\%.

In Figs.~\ref{fig:MchiMarg2} and \ref{fig:MchiMarg1}, we illustrate this effect, assuming that a Weyl fermion with fiducial $T_X=0.91$ K  and different values of $m_X$ is observed using different combinations of galaxy and CMB surveys. 
The broadening of the error bars is primarily driven by the LSS information and, as a consequence, the biggest shift in constraints is observed for datasets that are primarily or exclusively constrained by the galaxy surveys. 

We see in Fig.~\ref{fig:MchiMarg2} that the Planck constraint monotonically weakens with increasing fiducial relic mass. This can be explained by the decreasing effect of a relic on the radiation energy density $\rho_r$ which the CMB is primarily sensitive to. At low masses, the Planck dataset demonstrates an orthogonal relationship between the relic mass and degrees of freedom. Considering that at low masses,  changes in mass will modify the weak lensing signal of the CMB and produce no change in $\rho_r$ yet small changes in $g_X$ will produce directly proportional changes in $\rho_r$ we expect a nearly orthogonal relationship between these two parameters at low masses primarily governed by Eq.~\eqref{eq:Neff_relc=ic}, as the CMB signal is dominated by changes to $\rho_r$. However, as the fiducial relic mass is increased, and the relic effect on $\rho_r$ at recombination becomes smaller, the CMB becomes sensitive to the relic primarily through its effect on the weak lensing signal and the governing relationship changes to Eq.~\eqref{eq:omegaX} which is directly proportional to the product of $m_X$ and $g_X$. Thus, these two parameters are expected to develop an anti-correlation at high masses in the CMB dataset, which is indeed what we observe in Fig.~\ref{fig:MchiMarg2} at higher masses. 

Now we consider how the degeneracy direction in the $m_X-g_X$ plane varies at different relic masses for the LSS datasets. Here the scales affected by the relic, as governed by Eq.~\eqref{eq:kfs}, and by the magnitude of the effect, as determined by Eq.~\eqref{eq:omegaX}, control the effect on the LSS signal. At low masses, the contribution of the relic to $\omega_M$ is small and the relic will primarily affect the LSS signal through its free-streaming scale, which is independent of $g_X$. So at small relic masses, we expect $m_X$ and $g_X$ to be approximately orthogonal. As the fiducial relic mass is increased, the contribution of the relic to $P_M$ and hence to the LSS signal increases and is again proportional to the product of $m_X$ and $g_X$. So with increasing relic mass, we generally expect an anti-correlation to develop between the relic mass and degrees of freedom. We again see this to be the case in Fig. \ref{fig:MchiMarg2}. 

As discussed above, allowing the relic mass to vary modifies the constraints of the LSS and CMB datasets such that the accuracy of those constraints is generally less affected for lower mass relics. As the relic occupies a greater portion of $\Omega_M$, it becomes more important to simultaneously vary the relic mass and degrees of freedom. We emphasize that for a fixed relic abundance, there is a degeneracy between the relic parameters $m_X$, $T_X$, $g_X$ according to Eq.~\eqref{eq:omegaX}. This allows us to translate constraints on any two of these parameters into constraints on the third parameter. Where we have allowed the relic mass and degrees of freedom to vary, the resulting constraints can be translated to errors on the temperature. We also bring attention to the fact that marginalizing over the relic mass is only valid in the neighborhood of parameter space around each fiducial choice, and not over the entire parameter space of masses permitted.

%%%%%%%%%%%%%%%%%%%%%%%%%%%%%%%%%%%%%%%%%%%%%%%%%%%%
\section{Sampling of Full Model Posterior Forecasts}
\label{sec:AppB}

\begin{figure*}[h!]
    \centering
    \includegraphics[width = 1.0\linewidth]{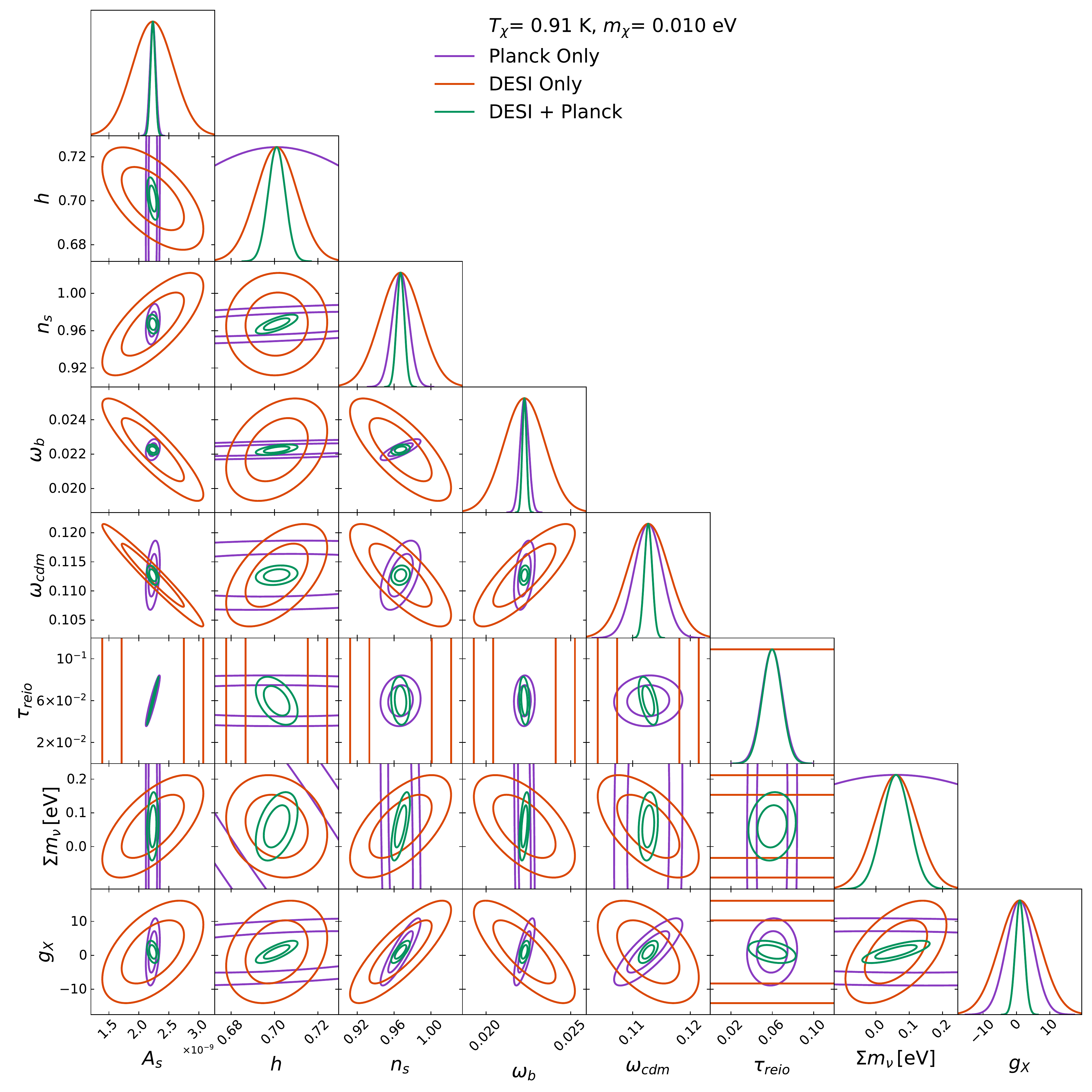}
    \caption{2-dimensional posterior distributions for parameter forecasts using {DESI} + \Planck, and each experiment individually. We assume here the presence of a Weyl fermion {\limr} ($g_X =2$) with $T_X$ = 0.91 K and $m_X$ = 0.01 eV. As shown, the complementarity between the two datasets results in marked improvement on the sensitivity to such a relic.} 
    \label{fig:triangle1}
\end{figure*}

Datasets with different parameter degeneracies can powerfully constrain parameters when combined. To illustrate this complementary effect between CMB and LSS surveys, we present a sampling of fully marginalized posteriors in Fig.~\ref{fig:triangle1} for a Weyl (neutrino-like) relic with temperature 0.91 K and mass $0.01$ eV. 
In each figure, we present constraints using only {DESI}  (red), only {\Planck} (violet), and the joint dataset (green). 

As in the case of the {\limr} parameter $g_X$ (number of degrees of freedom) discussed in the main text, the addition of LSS information to CMB data will generally break degeneracies between parameters. 
As an interesting example, we observe that the LSS provides a  measurement of $\omega_{\rm cdm}$ that is very close to orthogonal from the CMB one, breaking degeneracies with $A_s$, $n_s$ and $g_X$ for very light relic masses. 
DESI information also serves to set the measurements on $h$ and $\sum m_\nu$, which are poorly measured by \Planck as their effects on the CMB are degenerate.  
In turn, the LSS by itself is generally ineffective at measuring the other cosmological parameters, and provides no information on $\tau_{\rm reio}$. While, as illustrated in Fig.~\ref{fig:squareplot}, the degeneracies between $g_X$ and other parameters shift significantly between relics of different masses, those between the cosmological parameters themselves remain largely unchanged.

\captionsetup[figure]{justification=centerlast}

\end{document}